\documentstyle[preprint,eqsecnum,pre,aps]{revtex}
\tightenlines
\begin{document}
\draft
\title{
       Intermittency in coupled maps
      }
\author{Sang-Yoon Kim
       \footnote{Electronic address: sykim@cc.kangwon.ac.kr}
       }
\address{
Department of Physics \\ Kangwon National University \\
Chunchon, Kangwon-Do 200-701, Korea
        }

\maketitle
\begin{abstract}
Using a renormalization method, we study the critical behavior for
intermittency in two coupled one-dimensional (1D) maps. We find two 
fixed maps of the renormalization transformation. They all have 
common relevant eigenvalues associated with scaling of the control 
parameter of the uncoupled 1D map. However, the relevant ``coupling
eigenvalues'' associated with coupling perturbations vary depending
on the fixed maps. It is also found that the two fixed maps are 
associated with the critical behavior in the vicinity of a critical 
line segment. One fixed map with no relevant coupling eigenvalues 
governs the critical behavior at interior points of the critical 
line segment, while the other one with relevant coupling eigenvalues 
governs the critical behavior at both ends. The results of the two 
coupled 1D maps are also extended to many globally-coupled 1D maps, 
in which each 1D map is coupled to all the other ones with equal 
strength.
\end{abstract}
\pacs{PACS numbers : 05.45.+b, 03.20.+i, 05.70.Jk}
%
%

\narrowtext

\section{Introduction}
\label{sec:INT}

An intermittent transition to chaos in the 1D map occurs in the 
vicinity of a saddle-node bifurcation \cite{MP}. Intermittency just 
preceding a saddle-node bifurcation to a periodic attractor is 
characterized by the occurrence of intermittent alternations between 
laminar and turbulent behaviors. Scaling relations for the average 
duration of laminar behavior in the presence of noise have been 
first established \cite{EH} by considering a Langevin equation 
describing the map near the intermittency threshold and using 
Fokker-Plank techniques. The same scaling results have been later 
found \cite{HH} by using the same renormalization-group equation 
\cite{Feigenbaum} for period doubling with a mere change of boundary 
conditions appropriate to a saddle-node bifurcation.

Recently, efforts have been made to generalize the scaling results of 
period doubling for the 1D map to coupled 1D maps 
\cite{Kapral,Kuznet,Aranson,Kim1,Kim2,Kim3,Kim4}, which are used to 
simulate spatially extended systems with effectively many degrees of 
freedom \cite{Kaneko1}. It has been found that the critical scaling 
behaviors of period doubling for the coupled 1D maps are much richer 
than those for the uncoupled 1D map \cite{Kim1,Kim2,Kim3,Kim4}. These 
results for the abstract system of the coupled 1D maps are also 
confirmed in the real system of the coupled oscillators \cite{Kim5}. 
In a similar way, the scaling results of the higher period
$p$-tuplings $(p=,3,4,...)$ in the 1D map are also generalized to the 
coupled 1D maps \cite{Kim6}. Here we are interested in another route 
to chaos via intermittency in coupled 1D maps. Using a renormalization 
method, we extend the scaling results of intermittency for the 1D map 
to coupled 1D maps.

This paper is organized as follows. In Sec.~\ref{sec:TCM} we introduce
two coupled 1D maps and discuss their symmetry. Bifurcations
associated with stability of periodic orbits are also discussed there.
In Sec.~\ref{sec:RA} we employ the same renormalization method 
\cite{Kim3} developed for period doubling in coupled 1D maps and study 
the critical behavior for intermittency in two coupled 1D maps.
We thus find two fixed maps of the renormalization transformation.
They have the relevant eigenvalues associated with scaling of the 
control parameter of the uncoupled 1D map as common ones. However, 
the relevant ``coupling eigenvalues'' (CE's) associated with coupling 
perturbations vary depending on the fixed maps. These two fixed maps 
are also found to be associated with the critical behavior near a 
critical line segment in Sec.~\ref{sec:SB}. One fixed map with no 
relevant CE's governs the critical behavior at interior points of the 
critical line, while the other one with relevant CE's governs the 
critical behavior at both end points. We also study the critical 
behavior for intermittency in many globally-coupled 1D maps in 
Sec.~\ref{sec:MC}. Globally-coupled systems, in which each element is 
coupled to all the other ones with equal strength, appear naturally 
in broad branches of science \cite{Kaneko2}. The results of two 
coupled maps are extended to this kind of many globally-coupled maps. 
Finally, a summary is given in Sec.~\ref{sec:SUM}.

\section{Two coupled 1D maps}
\label{sec:TCM}

After briefly reviewing the intermittency in case of the 1D map, we 
introduce two coupled 1D maps and discuss their symmetry. Bifurcations
associated with stability of periodic orbits are also discussed.

We first recapitulate the intermittent transition to chaos
\cite{MP,EH,HH} in a 1D map with one control parameter $A$, 
$X_{t+1}=u(X_t)$ ($t$ denotes a discrete time). A pair of orbits with 
period $p$ appears via saddle-node bifurcation, as the control 
parameter $A$ exceeds a threshold value $A_c$. One periodic orbit is 
a stable attractor, while the other one is an unstable repeller. 
However, as $A$ decreases below $A_c$, the two periodic orbits 
disappear, and an intermittent chaotic attractor, characterized by the 
occurrence of intermittent alternations between laminar and turbulent 
behaviors, appears.

One can easily explain the intermittency geometrically as follows. The 
curve of the equation $Y=u^{(p)}(X)$ [$u^{(p)}$ is the $p$th iterate 
of $u$] has new $2p$ intersection points with the $Y=X$ line for 
$A>A_c$, which collapse into $p$ points tangent to the $Y=X$ line for 
$A=A_c$ (i.e., we have $p$ fixed points, $X^*_t = u^{(p)}(X^*_t)$, 
$t=1,\dots,p$ for $A=A_c$). However, as $A$ decreases below $A_c$
the curve no longer touches the $Y=X$ line so that a ``channel'' 
appears in the immediate vicinity of each point tangent to the $Y=X$ 
line at $A=A_c$. If one orbit point falls close to the entrance to 
one of the channels under repeated iterations of $u^{(p)}$, it would 
take many iterations to go through the channel, which corresponds to 
the ``laminar phase.'' After slow passage through the channel, the 
iterates of $u^{(p)}$ move wildly until they return to one of the 
channels. This corresponds to the ``turbulent phase.'' Thus the 
laminar and turbulent phases appear intermittently.

The nearer $A$ is to $A_c$, the longer the averaged laminar time.
To obtain the average duration of the laminar phase, consider the 
$p$th iterate $u^{(p)}(X)$ in the immediate vicinity of one of
the channels. Shifting the origin of coordinate $X$ to a fixed point
$X^*$ of $u^{(p)}(X)$ for $A=A_c$, we have
\begin{equation}
x_{t+1}=f(x_t) \equiv u^{(p)}(x_t+X^*) -X^* \approx
x_t + a |x_t|^z +\epsilon, \;\;z>1,
\label{eq:fm}
\end{equation}
where $x=X-X^*$, $a$ is a constant, and $\epsilon$ is a control
parameter proportional to $A-A_c$. Using a map $f$ of the form 
(\ref{eq:fm}), it has been found in \cite{HH} that the mean 
duration of the laminar phase $\bar l (\epsilon)$ varies as 
$\epsilon^{-(1-1/z)}$. Hence the tangency-order $z$ determines the 
universality classes, because $\bar l$ depends on the order $z$. In 
this paper, we consider only the analytic case of even order $z$ 
$(z=2,4,6,\dots)$.

We now consider a map $M$ consisting of two identical 1D maps
coupled symmetrically,
\begin{equation}
M: \left \{
        \begin{array}{l}
        X_{t+1}=W(X_t,Y_t)=u(X_t)+v(X_t,Y_t),\\
        Y_{t+1}=W(Y_t,X_t)=u(Y_t)+v(Y_t,X_t),
        \end{array}
   \right.
\label{eq:TCM}
\end{equation}
where $v$ is a coupling function obeying a condition,
\begin{equation}
v(X,X)=0\;\;{\rm for\;\;any}\;\;X.
\label{eq:CC}
\end{equation}

The map (\ref{eq:TCM}) is called a symmetric map because it has an 
exchange symmetry such that
\begin{equation}
{\sigma}^{-1}M{\sigma}({\bf Z})=M({\bf Z})\;\;{\rm for\;\;all\;\;}
{\bf Z},
\label{eq:ES}
\end{equation}
where ${\bf Z}=(X,Y)$, $\sigma$ is an exchange operator acting on 
$\bf Z$ such that $\sigma ({\bf Z})=(Y,X)$, and ${\sigma}^{-1}$ is 
its inverse. The set of all fixed points of $\sigma$ forms a 
synchronization line $Y=X$ in the state space. It follows from 
Eq.~(\ref{eq:ES}) that the exchange operator $\sigma$ commutes with 
the symmetric map $M$, i.e., $\sigma M = M \sigma$. Thus the 
synchronization line becomes invariant under $M$, i.e., if a point 
$\bf Z$ lies on the synchronization line, then its image $M (\bf Z)$ 
also lies on it. An orbit is called a(n) (in-phase) synchronous orbit 
if it lies on the synchronization line, i.e., it satisfies 
\begin{equation}
X_t=Y_t \equiv X^*_t \;\;{\rm for\;\;all\;\;}t.
\label{IO}
\end{equation}
Otherwise, it is called an (out-of-phase) asynchronous orbit. Here we 
study the intermittency associated with a saddle-node bifurcation to
a pair of synchronous periodic orbits, which can be easily found from 
the uncoupled 1D map, $X_{t+1}^* = u(X_t^*)$, because of the coupling 
condition (\ref{eq:CC}).

Stability of a synchronous orbit of period $p$ is determined from the
Jacobian matrix $J$ of $M^{(p)}$ ($p$th iterate of $M$), which is 
given by the $p$ product of the linearized map $DM$ of the map 
(\ref{eq:TCM}) along the orbit
\begin{eqnarray}
J &=& {\prod_{t=1}^{p}} DM(X^*_t,X^*_t) \nonumber \\
  &=& {\prod_{t=1}^{p}} \left (
  \begin{array}{cc}
  u'(X^*_t)-V(X^*_t) & V(X^*_t) \\
  V(X^*_t) & u'(X^*_t)-V(X^*_t)
  \end{array}
  \right ),
\label{eq:JM}
\end{eqnarray}
where $u'(x)=du(X)/dX$ and $V(X)= \partial v(X,Y)/ \partial 
Y |_{Y=X}$; hereafter $V(X)$ will be referred to as the ``reduced 
coupling function'' of $g(x,y)$. The eigenvalues of $J$, called the 
stability multipliers of the orbit, are gievn by
\begin{equation}
\lambda_1 = {\prod_{t}^{p}} u'(X^*_t),\;\;
\lambda_2 = {\prod_{t}^{p}} [u'(X^*_t) - 2V(X^*_t)].
\label{eq:SM}
\end{equation}
Note that $\lambda_1$ is just the stability multiplier for the case of 
the uncoupled 1D map and the coupling affects only $\lambda_2$.

A synchronous periodic orbit is stable when both multipliers lie 
inside the unit circle, i.e., $|\lambda_j| < 1$ for $j=1$ and $2$. 
Thus its stable region in the parameter plane is bounded by four 
bifurcation lines, i.e., those curves determined by the equations 
$\lambda_j=\pm1$ $(j=1,2)$. When a multiplier $\lambda_j$ increases
through $1$, the stable synchronous periodic orbit loses its stability
via saddle-node or pitchfork bifurcation. On the other hand, when
a multiplier $\lambda_j$ decreases through $-1$, it becomes unstable
via period-doubling bifurcation. (For more details on bifurcations,
refer to Ref.~\cite{Guckenheimer}.)

\section{Renormalization analysis of two coupled maps}
\label{sec:RA}

Here we are interested in intermittency just preceding a saddle-node
bifurcation. The intermittent transition to chaos occurs near a 
critical line segment, as will be seen in Sec.~\ref{sec:SB}. 
Using the same renormalization method \cite{Kim3,Kim4} for period 
doubling with a mere change of boundary conditions appropriate for a 
saddle-node bifurcation, we generalize the 1D scaling results for 
intermittency to the case of two coupled 1D maps. We thus find two 
fixed maps of the renormalization transformation and obtain their 
relevant eigenvalues. One fixed map is associated with the critical 
behavior inside the critical line, while the other one is associated 
with the critical behavior at both ends.

To study the intermittent transition to chaos near a saddle-node 
bifurcation to a pair of synchronous orbits of period $p$, consider 
the $p$th iterate $M^{(p)}$ of $M$ of Eq.~(\ref{eq:TCM}),
\begin{equation}
M^{(p)}: X_{t+1} = W^{(p)}(X_t,Y_t),\;Y_{t+1} = W^{(p)}(Y_t,X_t),
\label{eq:MM}
\end{equation}
where $W^{(p)}$ satisfies a recurrence relation
\begin{equation}
W^{(p)}(X,Y) = W (W^{(p-1)}(X,Y),W^{(p-1)}(Y,X)).
\end{equation}
The function $W^{(p)}$ can be decomposed into the uncoupled part
$u^{(p)}$ and the remaining coupling part, i.e.,
\begin{equation}
W^{(p)}(X,Y) = u^{(p)}(X) + [ W^{(p)}(X,Y) - u^{(p)}(X)].
\end{equation}
When the control parameter $A$ of the uncoupled 1D map is equal to 
the threshold value $A_c$ for the synchronous saddle-node 
bifurcation, we have $p$ synchronous fixed points of $M^{(p)}$ such 
that $Y^*_t=X^*_t=u^{(p)}(X^*_t)$ for $t=1,\dots,p$. Shifting the 
origin of coordinates $(X,Y)$ to one of the $p$ synchronous
fixed points $(X^*,Y^*)$ ($Y^*=X^*=u^{(p)}(X^*)$ for $A=A_c$), we have
\begin{equation}
T: \left \{
        \begin{array}{l}
        x_{t+1}=F(x_t,y_t)=f(x_t)+g(x_t,y_t),\\
        y_{t+1}=F(y_t,x_t)=f(y_t)+g(y_t,x_t),
        \end{array}
   \right.
\label{eq:TM}
\end{equation}
where
\begin{eqnarray}
f(x) &=& u^{(p)}(x+X^*) - X^*, \\
g(x,y) &=& W^{(p)}(x+X^*,y+Y^*) - u^{(p)}(x+X^*).
\end{eqnarray}
Since a synchronous saddle-node bifurcation occurs at the origin 
$(0,0)$ for $A=A_c$ in case of the map $T$, the uncoupled part $f$ for 
the critical case $A=A_c$ satisfies
\begin{equation}
f(0) = 0\;\;\;{\rm and}\;\;\;f'(0)=1.
\label{eq:BC}
\end{equation}
Note also that the coupling function $g(x,y)$ satisfies the coupling 
condition (\ref{eq:CC}), i.e.,
\begin{equation}
g(x,x)=0 \;\;\;{\rm for\;any}\;x.
\end{equation}

We employ the same renormalization transformation \cite{Kim3,Kim4} as 
in the period-doubling case with changed boundary conditions 
(\ref{eq:BC}). The renormalization transformation $\cal N$ for a 
coupled map $T$ consists of squaring $(T^2)$ and rescaling $(B)$ 
operators:
\begin{equation}
{\cal N}(T) \equiv B T^2 B^{-1}.
\label{eq:RT}
\end{equation}
Since we consider only synchronous orbits, the rescaling operator is
of the form,
\begin{equation}
B = \left(
         \begin{array} {cc}
           \alpha &\; 0 \\
           0 &\; \alpha
         \end{array}
    \right),
\end{equation}
where $\alpha$ is a rescaling factor.

Applying the renormalization operator ${\cal N}$ to the coupled map
(\ref{eq:TM}) $n$ times, we obtain the $n$-times renormalized map 
$T_n$ of the form,
\begin{equation}
 {T_n}:\left\{
       \begin{array}{l}
        x_{t+1} = {F_n}(x_t,y_t) = {f_n}(x_t) + {g_n}(x_t,y_t), \\
        y_{t+1} = {F_n}(y_t,x_t) = {f_n}(y_t) + {g_n}(y_t,x_t).
       \end{array}
     \right.
\label{eq:RTn}
\end{equation}
Here ${f_n}$ and ${g_n}$ are the uncoupled and coupling parts of the
$n$-times renormalized function $F_n$, respectively. They satisfy the
following recurrence equations \cite{Kim3,Kim4}:
\begin{eqnarray}
&f_{n+1}(x) = &
 \alpha {f_n}({f_n}({\frac {x} {\alpha}})), \label{eq:RUCFn} \\
&g_{n+1}(x,y) =& {\alpha} {f_n}({f_n}({\frac {x} {\alpha}}) + {g_n}
({\frac {x} {\alpha}},{\frac {y} {\alpha}})) \nonumber \\
&& +{\alpha} {g_n}({f_n}({\frac {x} {\alpha}}) + {g_n}({\frac {x} 
{\alpha}}, {\frac {y} {\alpha}}),{f_n}({\frac {y} {\alpha}}) 
\nonumber \\
&&\;\;\;\;\;\;\;\;\;\; +{g_n}({\frac {y} {\alpha}},{\frac {x} 
{\alpha}}))- {\alpha} {f_n}({f_n}({\frac {x} {\alpha}})).
\label{eq:RCFn}
\end{eqnarray}
Then Eqs.\ (\ref{eq:RUCFn}) and (\ref{eq:RCFn}) define a 
renormalization operator $\cal R$ of transforming a pair of functions, 
$(f,g)$;
\begin{equation}
 \left( \begin{array}{c}
                    {f_{n+1}} \\
                    {g_{n+1}}
                   \end{array}
           \right)
= {\cal R}  \left( \begin{array}{c}
                    {f_n} \\
                    {g_n}
                   \end{array}
           \right).
\label{eq:RORn}
\end{equation}

A critical map $T_c$ with the nonlinearity and coupling parameters
set to their critical values is attracted to a fixed map $T^*$ under
iterations of the renormalization transformation $\cal N$,
\begin{equation}
  {T^*}:\left\{
       \begin{array}{l}
        x_{t+1} = {F^*}(x_t,y_t) = {f^*}(x_t) + {g^*}(x_t,y_t), \\
        y_{t+1} = {F^*}(y_t,x_t) = {f^*}(y_t) + {g^*}(y_t,x_t) .
       \end{array}
     \right.
\label{eq:FM}
\end{equation}
Here $({f^*},{g^*})$ is a fixed point of the renormalization operator
$\cal R$, which satisfies $ (f^*,g^*) = {\cal R} (f^*,g^*) $.
Note that the equation for $f^*$ is just the fixed-point equation for
the intermittency with boundary conditions (\ref{eq:BC}) in the 
uncoupled 1D map. It has been found in \cite{HH} that
\begin{eqnarray}
f^*(x) &=& x [1-(z-1) a x^{z-1}]^{-1/(z-1)} \nonumber \\
&=& x+a x^z + {z \over 2}  a^2 x^{2z-1} + \cdots \;\; 
(a:{\rm arbitrary\;constant})
\label{eq:1Dfm}
\end{eqnarray}
is a fixed point of the transformation (\ref{eq:RUCFn}) with
\begin{equation}
\alpha = 2^{1/(z-1)}.
\end{equation}
[As mentioned in Sec.~\ref{sec:TCM}, we consider only the analytic
case of even order $z$ $(z=2,4,6, \dots).$] Consequently, only the 
equation for the coupling fixed function $g^*$ is left to be solved.

However, it is not easy to directly solve the equation for the 
coupling fixed function. We therefore introduce a tractable recurrence 
equation for a reduced coupling function of the coupling function 
$g(x,y)$, defined by
\begin{equation}
G(x) \equiv {\displaystyle \left.
{\frac {\partial g(x,y)} {\partial y}}\right|_{y=x}}.
\label{eq:RCFCT}
\end{equation}
Differentiating the recurrence equation (\ref{eq:RCFn}) for $g$ with
respect to $y$ and setting $y=x$, we have
\begin{eqnarray}
G_{n+1}(x) &=& [{f'_n}({f_n}({\frac {x} {\alpha}}))- 2G_{n}({f_n}
({\frac {x} {\alpha}}))] G_{n}({\frac {x} {\alpha}}) \nonumber \\
&& + G_{n}({f_n}({\frac {x} {\alpha}})) {f'_n}({\frac {x} 
{\alpha}}).
\label{eq:RRE}
\end{eqnarray}
Then Eqs.\  (\ref{eq:RUCFn}) and (\ref{eq:RRE}) define a 
``reduced renormalization operator'' $\tilde{\cal R}$ of transforming 
a pair of functions $(f,G)$ \cite{Kim3,Kim4}:
\begin{equation}
 \left( \begin{array}{c}
                    {f_{n+1}} \\
                    G_{n+1}
                   \end{array}
           \right)
= {\tilde{\cal R}}  \left( \begin{array}{c}
                    {f_n} \\
                    G_{n}
                   \end{array}
           \right).
\label{eq:RRO}
\end{equation}

We now look for a fixed point $({f^*},{G^*})$ of ${\tilde {\cal R}}$, 
which satisfies $ (f^*,G^*) = {\tilde {\cal R}} (f^*,G^*)$.
Here $G^*$ is just the reduced coupling fixed function of $g^*$ (i.e.,
${G^*}(x) = {\partial {g^*}(x,y)}/{\partial y}|_{y=x}$).
Using a series-expansion method, we find two solutions for $G^*$:
\begin{eqnarray}
G^*_I(x)&=& {1 \over 2} [1 + z a x^{z-1} + z (z-{1 \over 2})
            a^2 x^{2(z-1)} + \cdots], \label{eq:SS1} \\
G^*_{II}(x) &=& {1 \over 2} [b a x^{z-1} + b ({{3z-b} \over 2}-
                {1 \over 2}) a^2 x^{2(z-1)} + \cdots],
\label{eq:SS2}
\end{eqnarray}
where $a$ and $b$ are arbitrary constants.
Here we are able to sum the series in Eq.~(\ref{eq:SS1}) and obtain a
closed-form solution,
\begin{equation}
G^*_I(x) = {1 \over 2} {f^*}'(x).
\label{eq:CS}
\end{equation}
However, unfortunately we cannot sum the series in Eq.~(\ref{eq:SS2}),
except for the cases $b=0$ and $z$ where we obtain closed-form
solutions,
\begin{equation}
G^*_{II} (x) = \left \{
                  \begin{array}{c}
                  0 \;\;\;\;\;\;{\rm for}\;b=0, \\
                  {1 \over 2} [{f^*}'(x)-1]\;\;\;{\rm for}\;b=z.
                  \end{array}
          \right.
\end{equation}

Consider an infinitesimal perturbation $(h,\Phi)$ to a fixed point
$(f^*,G^*)$ of the reduced renormalization operator 
${\tilde{\cal R}}$. We then examine the evolution of a pair of 
functions $(f^* +h, G^* +\Phi)$ under $\tilde{\cal R}$. Linearizing 
$\tilde{\cal R}$ at its fixed point, we obtain a reduced linearized 
operator $\tilde{\cal L}$ transforming a pair of perturbations 
$(h,\Phi)$:
\widetext
\begin{equation}
 \left( \begin{array}{c}
                    {h_{n+1}} \\
                    {\Phi _{n+1}}
                   \end{array}
           \right)
= {\tilde{\cal L}}  \left( \begin{array}{c}
                     {h_n} \\
                     {\Phi_n}
                           \end{array}
                    \right)
= \left( \begin{array}{cc}
           {\tilde{\cal L}_1} & \;\;\; 0\\
           {\tilde{\cal L}_3} & \;\;\; {\tilde{\cal L}_2}
          \end{array}
           \right) \;
           \left( \begin{array}{c}
                    {h_n} \\
                    {\Phi_n}
                   \end{array}
              \right),
\end{equation}
where
\begin{eqnarray}
&{h_{n+1}}(x)&= [{{\tilde{\cal L}}_1} {h_n}](x) =
{\alpha}{f^*}'({f^*}(\frac{x}{\alpha})) {h_n}(\frac{x}{\alpha}) +
{\alpha} {h_n}({f^*}(\frac{x}{\alpha})), \label{eq:REh} \\
&{{\Phi}_{n+1}}(x) &= [{{\tilde{\cal L}}_2}{\Phi _n}](x)
      +[{{\tilde{\cal L}}_3}{h_n}](x), \label{eq:REPhi} \\
&[{{\tilde{\cal L}}_2}{\Phi_n}](x)&=
 [{f^*}'({f^*}(\frac{x}{\alpha}))-2{G^*}({f^*}(\frac{x}{\alpha}))]
{\Phi _n}(\frac{x}{\alpha})
+ [{f^*}'(\frac{x}{\alpha})-2{G^*}(\frac{x}{\alpha})] {\Phi _n}({f^*}
(\frac{x}{\alpha})),  \\
&[{{\tilde{\cal L}}_3}{h_n}](x)&= [{f^*}''({f^*}(\frac{x}{\alpha}))
{G^*}(\frac{x}{\alpha})-2{G^*}'({f^*} (\frac{x}{\alpha})){G^*}
(\frac{x}{\alpha}) +{G^*}'({f^*}(\frac{x}{\alpha}))
{f^*}'(\frac{x}{\alpha})] {h_n}(\frac{x}{\alpha}) \nonumber \\
 &&\;\;\; +{G^*}(\frac{x}{\alpha}){h'_n}({f^*}(\frac{x}{\alpha})) + 
 {G^*}({f^*}(\frac{x} {\alpha})){h'_n}(\frac{x}{\alpha}).
\end{eqnarray}
\narrowtext
Here the prime denotes a derivative. Note that, although $h_n$ couples 
to both $h_{n+1}$ and $\Phi_{n+1}$, $\Phi_n$ couples only to 
$\Phi_{n+1}$. From this reducibility of $\tilde{\cal L}$ into a 
semiblock form, it follows that to obtain the eigenvalues of 
$\tilde{\cal L}$ it is sufficient to solve the eigenvalue problems for 
${\tilde{\cal L}}_1$ and ${\tilde{\cal L}}_2$ separately. The 
eigenvalues of both ${\tilde{\cal L}}_1$ and ${\tilde{\cal L}}_2$ give 
the whole spectrum of $\tilde{\cal L}$.

A pair of perturbations $(h^*,\Phi^*)$ is called an eigenperturbation 
with eigenvalue $\lambda$, if
\begin{equation}
\lambda
\left( \begin{array}{c}
        h^*\\
        \Phi^*
        \end{array}
\right) =
{\tilde{\cal L}}
\left( \begin{array}{c}
        h^* \\
        \Phi^*
        \end{array}
\right),
\label{eq:EVE}
\end{equation}
that is,
\begin{eqnarray}
\lambda h^*(x) &=& [{\tilde{\cal L}_1}h^*](x), \label{eq:EVEh}\\
\lambda \Phi^*(x) &=&
[{\tilde{\cal L}_2} \Phi^*](x)+[{\tilde{\cal L}_3} h^*](x).
\end{eqnarray}
We first solve Eq.~(\ref{eq:EVEh}) to find eigenvalues of 
$\tilde{\cal L}_1$. Note that this is just the eigenvalue equation 
for the 1D map case. The complete spectrum of eigenvalues and the
corresponding eigenfunctions have been found in Refs.~\cite{HH}. The 
form of the eigenvalues is $\lambda_n = 2^{(z-n)/(z-1)}$ 
$(n=0,1,2,\dots)$. Hence the first $z$ eigenvalues with $n<z$ are 
relevant ones. The marginal one $\lambda_z$ is associated with the
arbitrary constant $a$ in $f^*(x)$, and all the other ones with 
$n>z$ are irrelevant. Although the eigenvalues $\lambda_n$'s of 
$\tilde{\cal L}_1$ are also eigenvalues of $\tilde{\cal L}$ as 
mentioned in the preceding paragraph, $(h^*,0)$ itself cannot be an 
eigenperturbation of $\tilde{\cal L}$ unless $\tilde{\cal L}_3$ is a 
null operator.

We next consider a perturbation of the form $(0,\Phi)$ having only 
the coupling part. If a reduced coupling perturbation $\Phi^*$ 
satisfies
\begin{eqnarray}
\lambda {\Phi^*}(x) &=& [{{\tilde{\cal L}}_2}{\Phi^*}](x) \nonumber \\
      &=&
 [{f^*}'({f^*}(\frac{x}{\alpha}))-2{G^*}({f^*}(\frac{x}{\alpha}))]
{\Phi^*}(\frac{x}{\alpha}) \nonumber \\
&&+ [{f^*}'(\frac{x}{\alpha})-2{G^*}(\frac{x}{\alpha})] {\Phi^*}({f^*}
(\frac{x}{\alpha})),
\label{eq:CE}
\end{eqnarray}
then it is called a reduced coupling eigenperturbation with coupling
eigenvalue (CE) $\lambda$. In this case $(0,\Phi^*)$ becomes an 
eigenperturbation of $\tilde{\cal L}$ with CE $\lambda$.

We first consider the case of the solution $G_I^*(x)={1 \over 2} 
{f^*}'(x)$. In this case the linearized operator $\tilde{\cal L}_2$ of 
Eq.~(\ref{eq:CE}) becomes a null operator because the right-hand side 
of the equation becomes zero. Hence there exist no relevant CE's 
associated with coupling perturbations, and consequently the fixed 
point $(f^*, G^*_I)$ has only relevant eigenvalues of 
$\tilde{\cal L}_1$ associated with scaling of the control parameter of 
the uncoupled 1D map. As will be seen in Sec.~\ref{sec:SB}, the 
intermittent transition to chaos occurs near a critical line 
segment. The critical behaviors near interior points of the critical 
line segment are essentially the same as those for the 1D case. In 
fact, a pair of critical functions $(f_c,G_c)$ at any interior point 
is attracted to the same fixed point $(f^*,G^*_I)$ under iterations 
of $\tilde{\cal R}$. Hence the fixed point $(f^*,G^*_I)$ governs the 
critical behaviors inside the critical line segment. (For details 
refer to the next section.)

Second, consider the case of the solution $G^*_{II}$ of 
Eq.~(\ref{eq:SS2}). Using a series-expansion method, we find the 
complete spectrum of CE's and the corresponding eigenfunctions. An 
eigenfunction $\Phi^*(x)$ can be expanded as follows:
\begin{equation}
\Phi^*(x) = \sum_{l=0}^{\infty} \,c_l \, x^l.
\end{equation}
Substituting the power series of $f^*(x)$, ${f^*}'(x)$, $G^*_{II}(x)$
and $\Phi^*(x)$ into the eigenvalue equation (\ref{eq:CE}), it has 
the structure
\begin{equation}
\lambda\, c_k\, = \,\sum_l \,{\cal M}_{kl}\, c_l,\;\;k,l=0,1,2,\dots~.
\end{equation}
Note that each $c_l$ $(l=0,1,2,\dots)$ in the right-hand side is 
involved only in the determination of coefficients of monomials $x^k$ 
with $k=l+m(z-1)$ $(m=0,1,2,\dots)$. Hence $\cal M$ becomes a lower 
triangular matrix. Its eigenvalues are therefore just diagonal 
elements:
\begin{equation}
\lambda_k = {\cal M}_{kk} = {2 \over {\alpha^k}} = 2^{(z-1-k)
\,/\,(z-1)},\;\;k=0,1,2,\dots~.
\end{equation}
The first $(z-1)$ eigenvalues $\lambda_k$'s for $0 \leq k \leq z-2$
are relevant ones. The marginal eigenvalue $\lambda_{z-1}$ is 
associated with the arbitrary constant $b$ in $G^*_{II}(x)$, and all 
the other eigenvalues for $k>z-1$ are irrelevant.

Each eigenfunction $\Phi^*_k(x)$ with CE $\lambda_k$ $(k=0,1,2,\dots)$
is of the form,
\begin{equation}
\Phi^*_k(x) = x^k [1 + (z-b+{k \over 2})a x^{z-1}+ \cdots ].
\label{eq:CEF}
\end{equation}
In case of the largest CE $\lambda_0=2$, we are able to sum the series
in Eq.~(\ref{eq:CEF}) for $b=0$ and $z$ and find $\Phi^*_0(x)$
in closed form:
\begin{equation}
\Phi^*_{0} (x) = \left \{
                  \begin{array}{c}
                  {f^*}'(x) \;\;{\rm for}\;b=0, \\
                  1\;\;\;{\rm for}\;b=z.
                  \end{array}
          \right.
\end{equation}
We can also sum the series in Eq.~(\ref{eq:CEF}) for all the 
irrelevant cases (i.e., the cases $k \geq z$) and find the 
closed-form eigenfunctions,
\begin{equation}
\Phi^*_k(x) = {1 \over {al}} [{f^*}'(x) - 2 G^*_{II}(x)]
              [{f^*}^l(x) - x^l],\;\;l=k-z+1=1,2,\dots~,
\end{equation}
which are associated with coordinate changes \cite{Kim7}.

As we shall see in the next section \ref{sec:SB}, the second reduced
coupling function $G^*_{II}(x)$ governs the critical behavior near 
both ends of a critical line segment, because a pair of critical
functions $(f_c,G_c)$ at each end point is attracted to the fixed 
point $(f^*,G^*_{II})$ under iterations of $\tilde{\cal R}$. (For 
details refer to the next section.)

\section{Scaling behavior near a critical line}
\label{sec:SB}

We choose the uncoupled 1D map in two coupled 1D maps $M$ of
Eq.~(\ref{eq:TCM}) as
\begin{equation}
u(X) = 1 - A X^2,
\label{eq:UCM}
\end{equation}
and consider a dissipative coupling case in which
the coupling function is given by
\begin{equation}
v(X,Y) = {c \over 2} [u(Y) - u(X)].
\label{eq:DC}
\end{equation}
Here $c$ is a coupling parameter. There exists a critical line segment
associated with intermittency on the synchronous saddle-node 
bifurcation line $A=A_c$ in the $c-A$ plane. Inside the critical line 
segment, the critical behaviors are governed by the fixed point 
$(f^*,G^*_I)$ with no relevant CE's, and hence they are essentially 
the same as those for the 1D case. On the other hand, the critical 
scalings near both ends are governed by the fixed point 
$(f^*,G^*_{II})$ with one relevant CE. As $c$ passes through both 
ends, the coupling leads to desynchronization of the interacting 
systems. Hence the synchronous attractor ceases to be an attractor 
and a new asynchronous attractor appears outside the critical line 
segment. Thus a transition from a synchronous to an asynchronous 
state occurs at both ends of the critical line segment.

As an example, we consider the saddle-node bifurcation to a pair of
synchronous orbits with period $p=3$ occurring for $A=A_c=1.75$.
To study the intermittency associated with this bifurcation, we first
consider the third iterate $M^{(3)}$ of $M$ [see eq.~(\ref{eq:MM})], 
and then shift the origin of coordinates $(X,Y)$ to one of the three 
synchronous fixed points $(X^*,Y^*)$ for $A=A_c$ $[Y^*=X^*=u^{(3)}
(X^*)]$. Thus we obtain a map $T$ of the form (\ref{eq:TM}), where the 
uncoupled and coupling parts $f$ and $g$ are given by
\begin{eqnarray}
f(x) &=& u^{(3)}(x+X^*) - X^*, \\
g(x,y) &=& W^{(3)} (x+X^*,y+Y^*) - u^{(3)}(x+X^*).
\end{eqnarray}
Near the region of the synchronous saddle-node bifurcation, $f(x)$ 
can be expanded about $x=0$ and $A=A_c$,
\begin{equation}
f(x) \approx  x + a x^2 + \epsilon,
\label{eq:up}
\end{equation}
where $a={1 \over 2} {\partial^2 f}/{\partial x^2}|_{x=0,\,A=A_c}$ and
      $\epsilon = {\partial f}/{\partial A}|_{x=0,\,A=A_c}\,(A-A_c)$.
Hence this corresponds to the most common case with the 
tangency-order $z=2$ [see Eq.~(\ref{eq:fm})]. The reduced coupling 
function $G(x)$ of $g(x,y)$ [defined in Eq.~(\ref{eq:RCFCT})] is also 
given by 
\begin{equation}
G(x) = {e \over 2}\, f'(x),\;\;\;e=c^3-3c^2+3c.
\label{eq:cer}
\end{equation}

Consider a pair of initial functions $(f_c,G)$ on the synchronous
saddle-node bifurcation line $A=A_c$, where $f_c(x)$ is just the 1D 
critical map and $G(x)={e \over 2} \,f'_c(x)$. By successive actions 
of the reduced renormalization transformation $\tilde {\cal R}$ of 
Eq.~(\ref{eq:RRO}) on $(f_c,G)$, we obtain
\begin{eqnarray}
&f_n&(x) = \alpha f_{n-1}(f_{n-1}({x \over \alpha})),\;\;
G_n(x)= {e_n \over 2}\,f'_n(x), \\
&e_n& = 2e_{n-1} -  e^2_{n-1}, \;\; (n=0,1,2,\dots),
\label{eq:red}
\end{eqnarray}
where the rescaling factor for $z=2$ is $\alpha=2$, $f_0(x) = f_c(x)$,
$G_0(x)=G(x)$, and $e_0=e$. Here $f_n$ converge to the 1D fixed 
function $f^*(x)$ of Eq.~(\ref{eq:1Dfm}) with $z=2$ as 
$n \rightarrow \infty$.

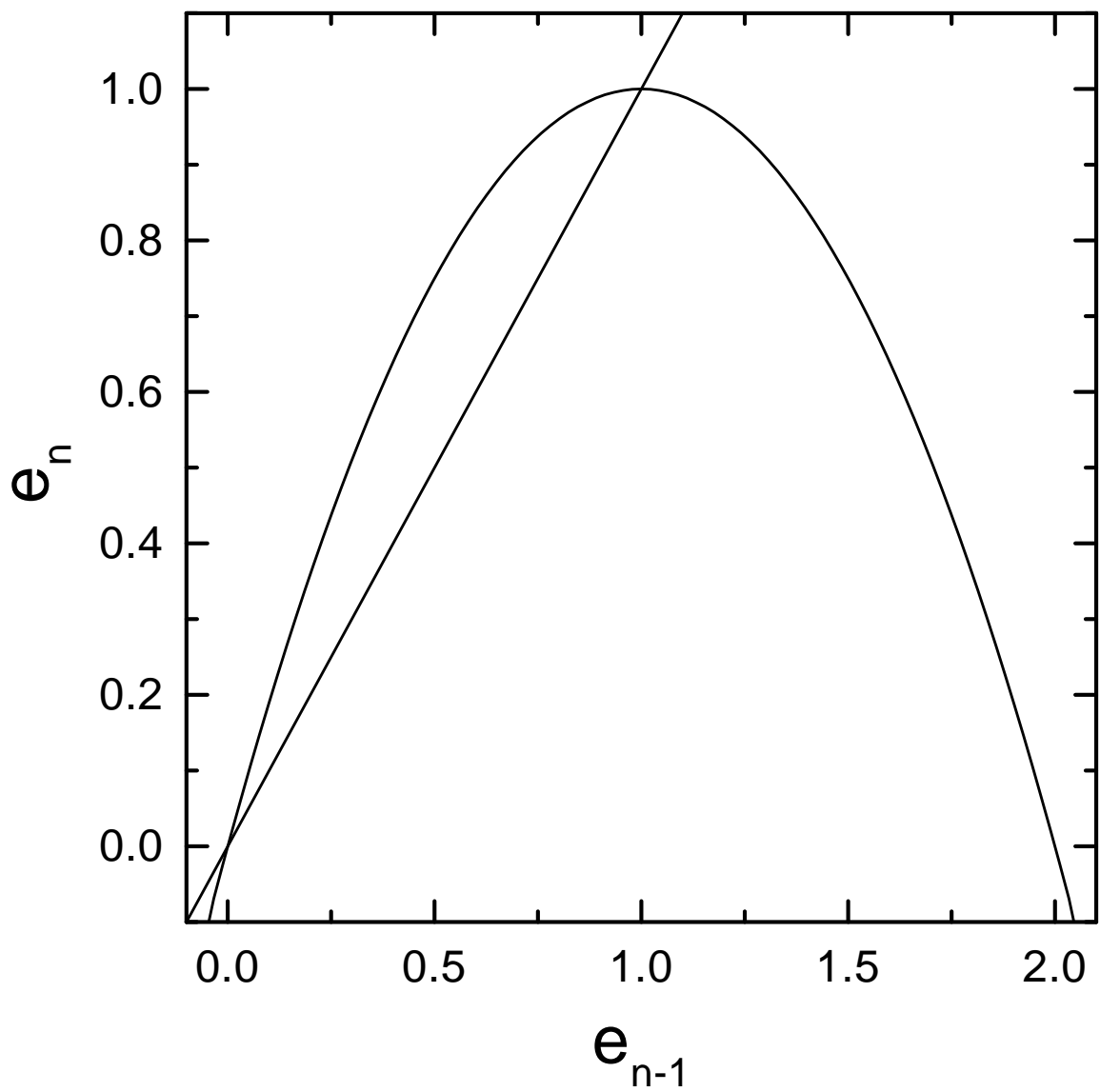
\begin{figure}
\caption{
Plots of the curve $e_n = 2 e_{n-1} - {e_{n-1}^2}$ and the straight 
line $e_n = e_{n-1}$.
}
\label{fig:RT}
\end{figure}

Figure \ref{fig:RT} shows a plot of the curve determined by 
Eq.~(\ref{eq:red}). Two intersection points between this curve and 
the straight line $e_n=e_{n-1}$ are just the fixed points $e^*$ of the 
recurrence relation (\ref{eq:red}) for $e$,
\begin{equation}
e^*=0,\,1.
\end{equation}
Stability of each fixed point $e^*$ is determined by its stability
multiplier $\lambda$ $[=de_n/de_{n-1}|_{e^*}]$.
The fixed point at $e^*=1$ is superstable $(\lambda=0)$, while
the other one at $e^*=0$ is unstable $(\lambda=2)$.
The basin of attraction to the superstable fixed point $e^*=1$
is the open interval $(0,2)$. That is, any initial $e$ inside the
interval $0<e<2$ converges to $e^*=1$ under successive
iterations of the transformation (\ref{eq:red}). The left end of the
interval is just the unstable fixed point $e^*=0$, which is also the
image of the right end point under the recurrence equation 
(\ref{eq:red}). All the other points outside the interval diverge to 
the minus infinity under iterations of the transformation 
(\ref{eq:red}).

It follows from the relation $e=e(c)$ in Eq.~(\ref{eq:cer}) that there
exists a critical line segment joining two end points $c_l=0$ and
$c_r=2$ on the synchronous saddle-node bifurcation line $A=A_c$ in the
$c-A$ plane. Inside the critical line segment, any initial $G(x)$ is
attracted to the first reduced coupling fixed function
$G^*_I(x)={1 \over 2} {f^*}'(x)$ under iterations of 
$\tilde {\cal R}$, while $G(x)$'s at both ends are attracted to the 
second reduced coupling fixed function $G^*_{II}(x)=0$ with $b=0$.

Figure \ref{fig:PD} shows a phase diagram for the case of a 
dissipative coupling (\ref{eq:DC}). The diagram is obtained from 
calculation of Lyapunov exponents. For the case of a synchronous 
orbit, its two Lyapunov exponents are given by
\begin{eqnarray}
\sigma_\|(A) &=& \lim_{m \rightarrow \infty} {1 \over m} 
{\sum_{t=0}^{m-1}} \ln{|u'(X_t)|}, 
\label{eq:sigmal} \\
\sigma_\bot (A,c) &=& \lim_{m \rightarrow \infty} {1 \over m} 
{\sum_{t=0}^{m-1}} \ln{|(1-c) u'(X_t)|}= \sigma_\|(A) + \ln{|1-c|}.
\label{eq:sigmac}
\end{eqnarray}
Here $\sigma_\| (\sigma_\bot)$ denotes the mean exponential rate of 
divergence of nearby orbits along (across) the synchronization line 
$Y=X$. Hereafter, $\sigma_\|$ $(\sigma_\bot)$ will be referred to as
tangential and transversal Lyapunov exponents, respectively. Note also
that $\sigma_\|$ is just the Lyapunoiv exponent for the 1D case, and
the coupling affects only $\sigma_\bot$.

The data points on the $\sigma_\bot = 0$ curve are denoted by solid 
circles in Fig.~\ref{fig:PD}. A synchronous orbit on the 
synchronization line becomes a synchronous attractor with 
$\sigma_\bot <0$ inside the $\sigma_\bot =0$ curve. 
The type of this synchronous attractor is determined according to the
sign of $\sigma_\|$. A synchronous period-3 orbit with $\sigma_\|<0$ 
becomes a synchronous periodic attractor above the critical line 
segment, while a synchronous chaotic attractor with $\sigma_\| > 0$ 
exists below the critical line segment. The periodic and chaotic 
parts in the phase diagram are denoted by $P$ and $C$, respectively. 
There exists also a synchronous period-3 attractor with 
$\sigma_\| =0$ on the critical line segment between the two parts.

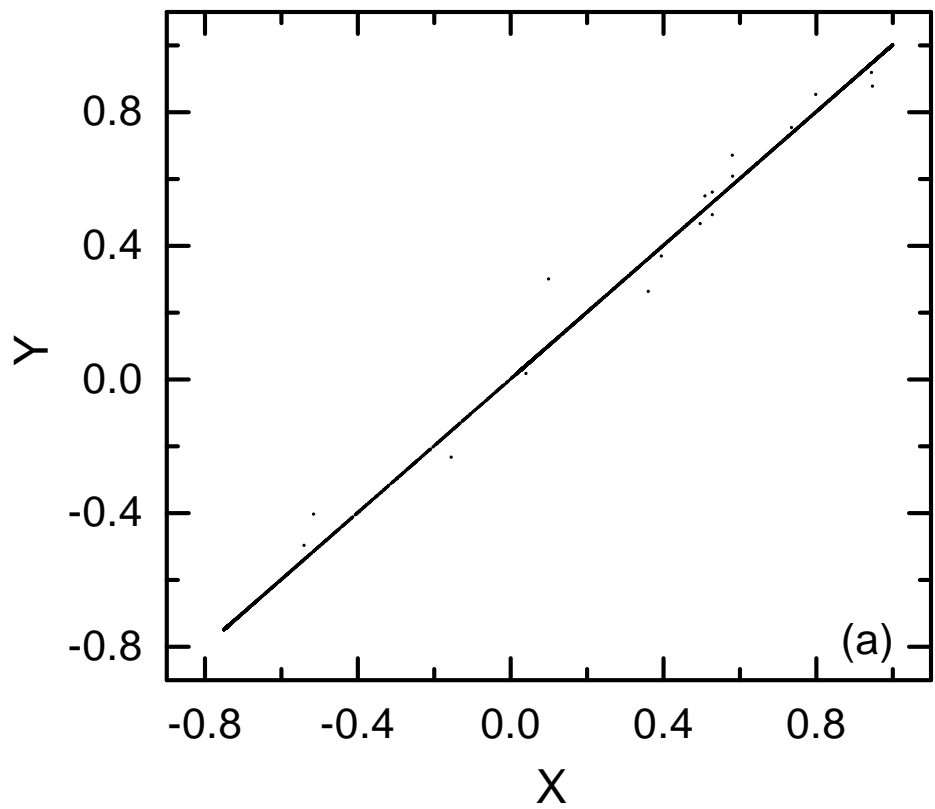
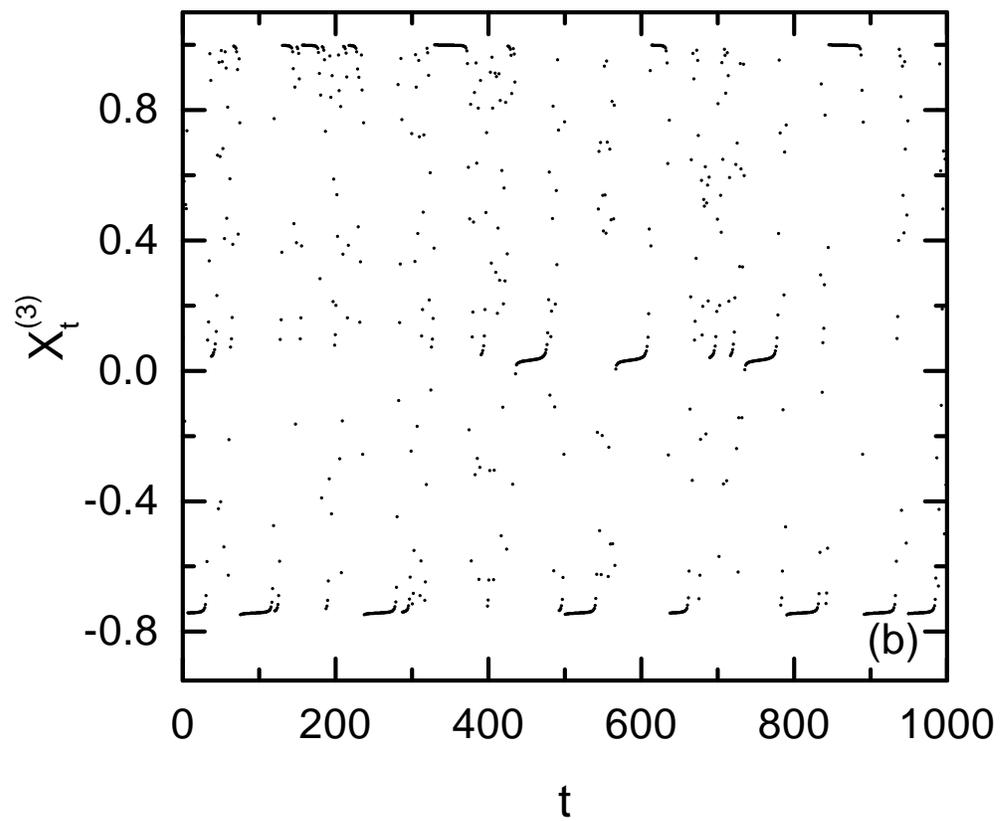
\begin{figure}
\caption{
Iterates of the dissipatively-coupled map $M$ of Eq.~(2.2) with the
uncoupled 1D map (4.1) and the coupling (4.2) for $\epsilon_A$
$(=A_c -A)=10^{-4}$ and $c=0.5$; (a) an orbit with initial point 
$(0.1,0.3)$ is attracted to a synchronous chaotic attractor on the 
synchronization line $Y=X$ after exhibiting a short transient 
behavior, and (b) the plot of the $X$-component $X_t^{(3)}$ of the 
third iterate $M^{(3)}$ of $M$ versus a discrete time $t$ shows 
intermittent alternations between laminar and turbulent behaviors.
}
\label{fig:1DIM}
\end{figure}

Consider a transition to chaos near an interior point with
$c_l < c < c_r$ of the critical line segment. Here we fix the value 
of the coupling parameter $c$ and vary the control parameter 
$\epsilon_A$ $(\equiv A_c -A)$. For $\epsilon_A <0$ there exists a 
synchronous period-3 attractor on the synchronization line $Y=X$. 
However, as $\epsilon_A$ is increased from zero, the periodic 
attractor disappears and a new chaotic attractor appears on the
synchronization line. As an example, see the figure \ref{fig:1DIM}(a) 
which shows a synchronous chaotic attractor for the case $\epsilon_A
= 10^{-4}$. The motion on this synchronous chaotic attractor is 
characterized by the occurrence of intermittent alternations between
laminar and turbulent behaviors on the synchronization line, as shown
in Fig.~\ref{fig:1DIM}(b). This is just the intermittency occurring 
in the uncoupled 1D map, because the motion on the synchronization
line is the same as that for the uncoupled 1D case. Thus, a 
``1D-like'' intermittent transition to chaos occurs near interior 
points of the critical line segment.

The scaling relations of the mean duration of the laminar phase 
$\bar l$ and the tangential Lyapunov exponent $\sigma_\|$ for a 
synchronous chaotic attractor are obtained from the leading relevant 
eigenvalues $\delta_1$ $(=4)$ of the fixed point $(f^*,G^*_I)$ of
$\tilde {\cal R}$ with no relevant CE's, like the 1D case \cite{HH}.
A map with non-zero $\epsilon$ near an interior point of the critical 
line segment is transformed into a new map of the same form, but with 
a new parameter $\epsilon'$ under a renormalization transformation. 
Here the control parameter scales as
\begin{equation}
\epsilon_A' = \delta_1 \epsilon_A = 2^2 \epsilon_A.
\end{equation}
Then the mean duration $\bar l$ and the tangential Lyapunov exponent 
$\sigma_\|$ satisfy the homogeneity relations,
\begin{equation}
{\bar l}({\epsilon_A'}) = {1 \over 2} {\bar l}(\epsilon_A),\;\;
{\sigma_\|}({\epsilon_A'}) = 2 \sigma_\| (\epsilon_A),
\end{equation}
which lead to the scaling relations,
\begin{equation}
{\bar l}(\epsilon_A) \sim \epsilon_A^{-\mu},\;\;{\sigma_\|}
(\epsilon_A) \sim {\epsilon_A^\mu},
\end{equation}
with exponent
\begin{equation}
\mu = \log2 / \log \delta_1 = 0.5.
\end{equation}

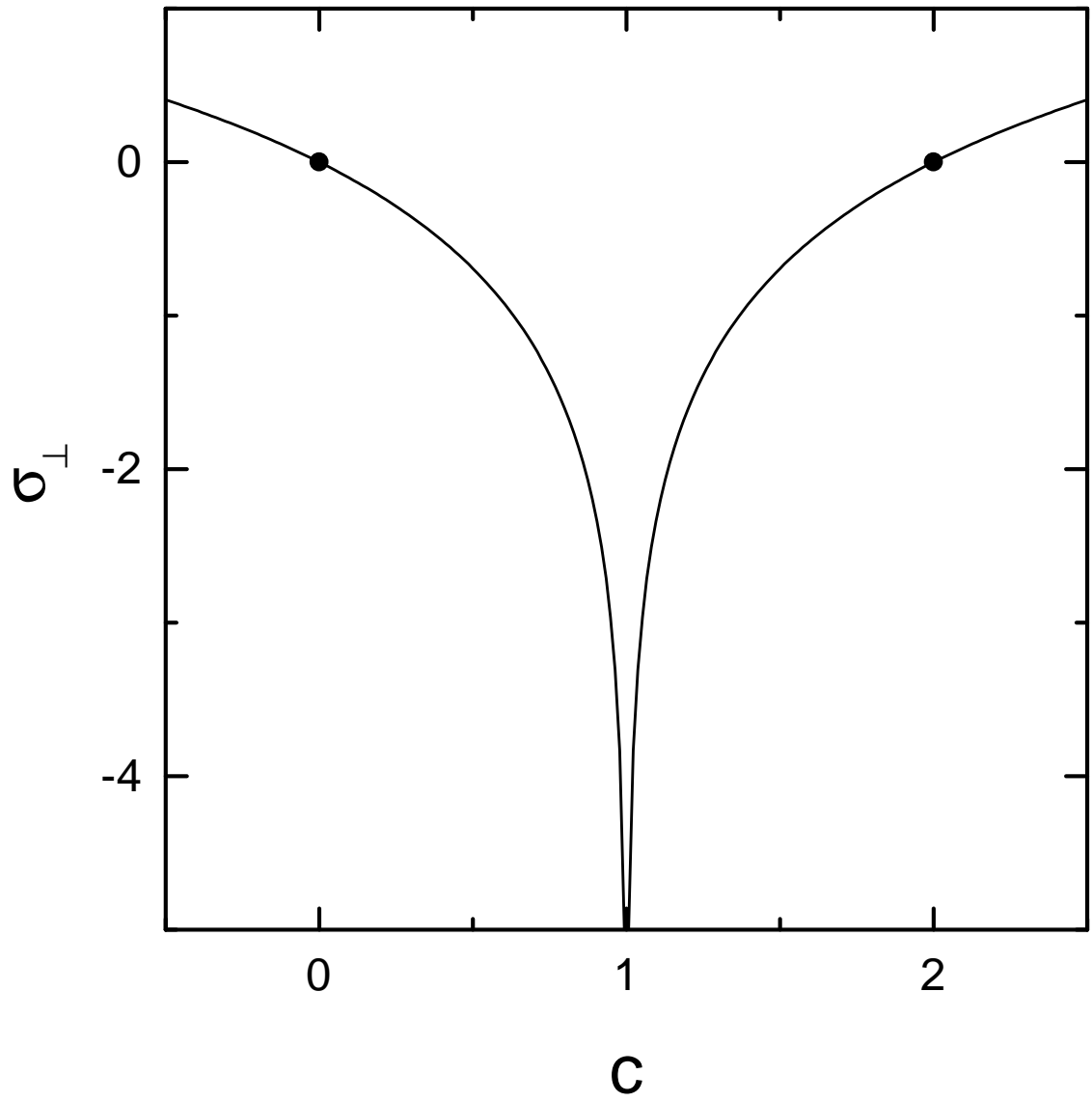
\begin{figure}
\caption{
Plot of $\sigma_\bot$ $(=\ln|1-c|)$ versus c for $A=A_c$. The values
of $\sigma_\bot$ at both ends of the critical line segment are zero, 
which are denoted by solid circles.
}
\label{fig:TLEXP}
\end{figure}

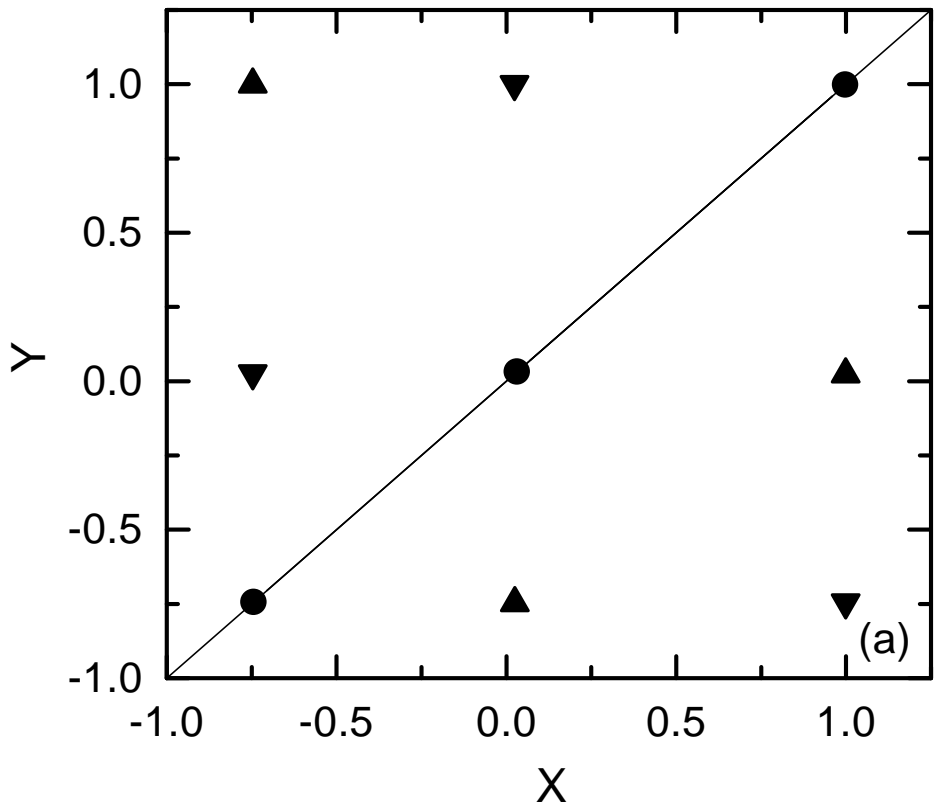
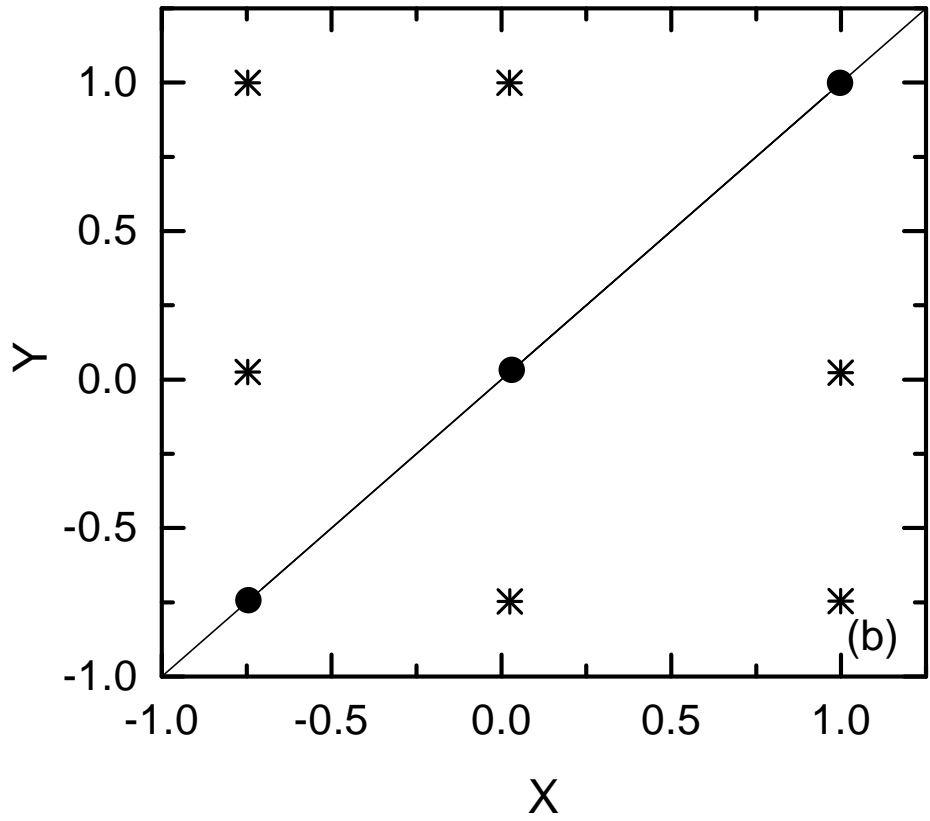
\begin{figure}
\caption{
Transition from a synchronous to an asynchronous state for $A=A_c$. A
synchronous period-3 orbit denoted by circles is no longer an 
attractor outside the critical line segment, and new asynchronous 
attractors exist for (a) $c=-0.0001$ and (b) $c=2.0001$. We denote a 
pair of asynchronous period-3 attractors by uptriangles and 
downtriangles in case of (a), while the asynchronous attrractor with
period 6 is denoted by ``stars'' in case of (b).
}
\label{fig:DSYN}
\end{figure}

The 1D-like intermittent transition to chaos ends at both ends $c_l$ 
and $c_r$ of the critical line segment. We fix the value of the 
control parameter $A=A_c(=1.75)$ and study the critical behaviors near 
the two end points by varying the coupling parameter $c$. Inside the 
critical line segment $(c_l < c < c_r)$, a synchronous period-3 
attractor exists on the synchronization line. However, as the coupling 
parameter $c$ passes through $c_l$ or $c_r$, the transversal Lyapunov 
exponent $\sigma_\bot$ of the synchronous periodic orbit increases 
from zero, as shown in Fig.~\ref{fig:TLEXP}, and hence the coupling 
leads to desynchronization of the interacting systems. Thus the 
synchronous period-3 orbit ceases to be an attractor outside the 
critical line segment, and a new (out-of-phase) asynchronous 
attractor appears. This is illustrated in Fig.~\ref{fig:DSYN}.
There exist an asynchronous attractor with period 3 denoted by 
uptriangles for $c=-0.0001$, while an asynchronous period-6
attractor denoted by ``stars'' exists for $c=2.0001$.
Since the asynchronous period-3 attractor is not a symmetric one 
under the exchange operator $\sigma$ of Eq.~(2.4), there exists also 
its conjugate asynchronous period-3 attractor denoted by
downtriangles. But the asynchronous period-6 attractor is a symmetric 
one under the exchange operator $\sigma$.

The critical behaviors near both ends $c_l$ and $c_r$ are the same.
Since the transversal Lyapunov exponent $\sigma_\bot$ is equal to 
$\ln |1-c|$ for $A=A_c$ [see Eq.~(\ref{eq:sigmac})], it varies 
linearly with respect to $c$ near both ends, (i.e., 
$\sigma_\bot \sim \epsilon_c$, $\epsilon_c \equiv c_l - c\;{\rm or}\; 
c-c_r$). The scaling behavior of $\sigma_\bot$ is obtained from the 
relevant CE $\delta_2$ $(=2)$ of the fixed point $(f^*,G^*_{II})$ of 
$\tilde {\cal R}$ as follows. Consider a map with non-zero 
$\epsilon_c$, (but with $\epsilon_A=0$) near both ends.
It is then transformed into a new one of the same form, but with a
renormalized parameter $\epsilon_c'$ under a renormalization 
transformation. Here the parameter $\epsilon_c$ obeys a scaling law,
\begin{equation}
\epsilon_c' = \delta_2\, \epsilon_c = 2 \epsilon_c.
\end{equation}
Then the transversal Lyapunov exponent $\sigma_\bot$ satisfies the 
homogeneity relation,
\begin{equation}
\sigma_c(\epsilon_c') = 2 \sigma_c (\epsilon_c).
\end{equation}
This leads to the scaling relation,
\begin{equation}
\sigma_c (\epsilon_c) \sim \epsilon_c^\nu,
\end{equation}
with exponent
\begin{equation}
\nu = \log 2 / \log {\delta_2} = 1.
\end{equation}

Like the case of interior points of the critical line segment, the 
scaling behavior of $\sigma_\| (\epsilon_A)$ for $c=c_l$ or $c_r$ is 
obtained from the relevant eigenvalue $\delta_1$ $(=4)$ of the fixed 
point $(f^*,G^*_{II})$, and hence it also satisfies the scaling 
relation (\ref{eq:sigmal}). The scaling behaviors of the Lyapunov 
exponents $\sigma_\|$ and $\sigma_\bot$ near both ends are thus 
determined from two relevant eigenvalues $\delta_1$ and $\delta_2$ of 
the fixed point $(f^*,G^*_{II})$.

\section{Extension to many globally-coupled maps}
\label{sec:MC}

Recently, much attention has been paid to dynamical systems with
many nonlinear elements and a global coupling, in which each element
is coupled to all the other elements with equal strength.
Globally-coupled systems as the extreme limit of long-range
couplings are seen in broad branches of science \cite{Kaneko2}.
For example, coupled nonlinear oscillators with a global coupling
frequently occur in charge density waves \cite{Als}, Josepson 
junction arrays \cite{Wiesenfeld}, and p-n junction arrays
\cite{Fabiny}. This kind of dynamical systems with a global coupling 
can be also regarded as mean-field versions of dynamical systems with 
local short-range couplings.

Here we study the critical behavior for intermittency in many-coupled
1D maps with a global coupling. The results of two coupled maps are 
straightforwardly extended to the many globally-coupled maps.

\subsection{Many globally-coupled 1D maps}
\label{sub:MCM}

Consider an $N$-coupled map with a periodic boundary condition:
\begin{eqnarray}
M: X_{t+1}(m) &=& W ({\sigma}^{m-1} {\bf X}_t ) \nonumber \\
 &=& W (X_t(m),X_t(m+1),\dots,X_t(m-1)), \nonumber \\
 && \;\;\;\;\;\;\;\;\;\;\;\;\;\;\;\;\;\;\;\;\;\;\;\;m=1,\dots,N,
\label{eq:mcm}
\end{eqnarray}
where the number of constituent elements $N$ is a positive integer 
larger than or equal to 2, $X_t(m)$ is the state of the $m$th element 
at a lattice point $m$ and at a discrete time $t$, 
${\bf X} = (X(1),X(2),\dots,X(N))$, $\sigma$ is the cyclic permutation 
of the elements of ${\bf X}$, i.e., $\sigma {\bf X} =(X(2), 
\dots,X(N),X(1))$, $\sigma^{m-1}$ means $(m-1)$ applications of 
$\sigma$. The periodic condition imposes $X_t(m) = X_t(m+N)$ for all 
$m$. Like the two-coupled case $(N=2)$, the function $W$ consists of 
two parts:
\begin{equation}
W({\bf X}) = u(X(1)) + v({\bf X}),
\end{equation}
where $u$ is an uncoupled 1D map and $v$ is a coupling function 
obeying the condition,
\begin{equation}
v(X,\dots,X) = 0 \;\;\;{\rm for\;any\;}X.
\label{eq:ccd}
\end{equation}
Thus the $N$-coupled map (\ref{eq:mcm}) becomes:
\begin{eqnarray}
M: X_{t+1}(m) &=& u(X_t(m)) \nonumber \\
   && + v(X_t(m),X_t(m+1),\dots,X_t(m-1)), \nonumber \\
  &&   \;\;\;\;\;\;\;\;\;\;\;\;\;\;\;\;\;\;\;\;\;\;m=1,\dots,N.
\label{eq:mcm2}
\end{eqnarray}

Here we study many-coupled 1D maps with a global coupling.
In the extreme long-range case of global coupling, the coupling 
function $v$ is of the form,
\begin{equation}
v(X(1),\dots,X(N)) = { c \over N} \sum_{m=1}^{N} [r(X(m))-r(X(1))]
\;\;{\rm for}\;N \geq 2,
\label{eq:gc}
\end{equation}
where $r(X)$ is a function of one variable.
Note that each 1D map is coupled to all the other ones with equal
coupling strength ${c/N}$ inversely proportional to the number of 
degrees of freedom $N$. Hereafter, $c$ will be called the coupling 
parameter.

The $N$-coupled map $M$ is called a symmetric map, because it has a 
cyclic permutation symmetry such that
\begin{equation}
\sigma^{-1} M \sigma ({\bf X}) = M ({\bf X})\;\;{\rm for\;all\;}
{\bf X},
\label{eq:CP}
\end{equation}
where $\sigma^{-1}$ is the inverse of $\sigma$.
The set of all fixed points of $\sigma$ forms a synchronization line 
in the $N$-dimensional state space, on which
\begin{equation}
X(1)=\cdots=X(N).
\end{equation}
It follows from Eq.~(\ref{eq:CP}) that the cyclic permutation $\sigma$
commutes with the symmetric map $M$, i.e., $\sigma M = M \sigma$. 
Hence the synchronization line becomes invariant under $M$. An orbit 
is called a(n) (in-phase) synchronous orbit if it lies on the 
synchronization line, i.e., it satisfies
\begin{equation}
X_t(1) = \cdots = X_t(N) \equiv X^*_t\;\;{\rm for \; all\;}t.
\label{eq:sl}
\end{equation}
Otherwise, it is called an (out-of-phase) asynchronous orbit. Here we
study the intermittency associated with a synchronous saddle-node 
bifurcation. Note also that synchronous orbits can be easily found 
from the uncoupled 1D map, $X^*_{t+1} = u(X^*_t)$, because of the 
coupling condition (\ref{eq:ccd}).

\subsection{Lyapunov exponents of synchronous orbits and the critical
            behavior}
\label{sub:LE}

Consider an orbit $\{ {\bf X}_t \} \equiv \{ X_t(m),\, m=1,\dots,N \}$ 
in many coupled maps (\ref{eq:mcm2}).
Stability analysis of the orbit can be easily carried out by Fourier
transforming with respect to the discrete space $\{ m \}$. The 
discrete spatial Fourier transform of the orbit is:
\begin{eqnarray}
{\cal F} [ X_t(m) ] \equiv { 1 \over N }  \sum_{m=1}^{N}
e^{-2 \pi i m j / N} \, X_t(m) = \xi_t(j), \nonumber \\
\;\;\;\;\;\;\;\;\;\;\;\;\;\;\;\;\;\;\;\;\;\; j=0,1,\dots,N-1.
\end{eqnarray}
The Fourier transform $\xi_t(j)$ satisfies $\xi_t^*(j) = \xi_t(N-j)$
($*$ denotes complex conjugate), and the wavelength of a mode with 
index $j$ is $N \over j$ for $j \leq {N \over 2}$ and 
$N \over {N-j}$ for $j > {N \over 2}$. Here $\xi_t(0)$ corresponds to
the synchronous (Fourier) mode of the orbit, while all the other
$\xi_t(j)$'s with nonzero indices $j$ correspond to the asynchronous
(Fourier) modes.

To determine the stability of a synchronous orbit
$[X_t(1) = \cdots =X_t (N) \equiv X^*_t$ for all $t$], we consider an
infinitesimal perturbation $\{ \Delta X_t(m) \}$ to the synchronous 
orbit, i.e., $X_t(m) = X^*_t + \Delta X_t(m)$ for $m=1,\dots,N$.
Linearizing the $N$-coupled map (\ref{eq:mcm2}) at the synchronous 
orbit, we obtain:
\begin{equation}
\Delta X_{t+1}(m) = u'(X^*_t) \Delta X_t(m) +
   {\sum_{l=1}^N} V^{(l)}(X^*_t) \Delta X_t(l+m-1),
\label{eq:LE1}
\end{equation}
where
\begin{eqnarray}
u'(X) = {du \over dx}, \;\;  V^{(l)}(X) &\equiv& \left.
{{\partial v(\sigma^{(m-1)}
{\bf X})} \over {\partial X_{l+m-1}} } \right |_{X(1)=\cdots=X(N)=X}
\nonumber \\
&=& \left. {{\partial v({\bf X})} \over {\partial X_l} }
\right |_{X(1)=\cdots=X(N)=X}.
\end{eqnarray}
Hereafter, the functions $V^{(l)}$'s will be called ``reduced coupling
functions'' of $v({\bf X})$.

Let $\delta \xi_t(j)$ be the Fourier transform of $\Delta X_t(m)$, 
i.e.,
\begin{eqnarray}
\delta \xi_t(j) = {\cal F} [\Delta X_t(m) ] = { 1 \over N }
\sum_{m=1}^{N} e^{-2 \pi i m j / N} \, \Delta X_t(m), \nonumber \\
\;\;\;\;\;\;\;\;\;\;\;\;\;\;\;\;\;\;\;\;\;\; j=0,1,\dots,N-1.
\end{eqnarray}
Here $\delta \xi_t(0)$ is the synchronous-mode perturbation along the
synchronization line, while all the other ones $\delta \xi_t(j)$'s
with nonzero indicies $j$ are the asynchronous-mode perturbations
across the synchronization line. Then the Fourier transform of 
Eq.~(\ref{eq:LE1}) becomes:
\begin{eqnarray}
\delta \xi_{t+1}(j) &=& [u'(X^*_t) +
   {\sum_{l=1}^N} V^{(l)}(X^*_t) e^{2\pi i (l-1) j /N} ]\, \delta 
    \xi_t(j),
   \nonumber \\
&&\;\;\;\;\;\;\;\;\;\;\;\;\;\;\;\;\;\;\;\;\;\; j=0,1,\dots,N-1.
\label{eq:LE2}
\end{eqnarray}
Note that all the modes $\xi_t(j)$'s become decoupled for the 
synchronous orbit.

The Lyapunov exponent $\lambda_j$ of a synchronous orbit, 
characterizing the average exponential rate of divergence of 
the $j$th-mode perturbation, is given by
\begin{equation}
\sigma_j = \displaystyle {\lim _{m \rightarrow \infty}} {1 \over m}
{\sum _{t=0}^{m-1}} \ln|
u'(X^*_{t}) +   {\sum_{l=1}^N} V^{(l)}(X^*_{t}) 
e^{2\pi i (l-1) j /N}|.
\label{eq:lexp}
\end{equation}
If the Lyapunov exponent $\sigma_j$ is negative or zero, then the 
synchronous orbit is stable against the $j$th-mode perturbation; 
otherwise it is unstable.

In case of a synchronous periodic orbit with period $p$, the Lyapunov
exponent $\sigma_j$ in Eq.~(\ref{eq:lexp}) becomes:
\begin{equation}
\sigma_j = {1 \over p} \ln|\lambda_j|,
\end{equation}
where $\lambda_j$, called the stability multiplier of the synchronous
orbit, is given by
\begin{equation}
\lambda_j =
{\prod_{t=0}^{p-1} }
[u'(X^*_{t}) +   {\sum_{l=1}^N} V^{(l)}(X^*_{t}) e^{2\pi i (l-1) 
 j /N} ].
\label{eq:sm}
\end{equation}
The synchronous periodic orbit is stable against the $j$th-mode 
perturbation when $\lambda_j$ lies inside the unit circle, i.e.,
$|\lambda_j| < 1$. When the stability multiplier $\lambda_j$ increases
through $1$, the synchronous periodic orbit loses its stability via
saddle-node or pitchfork bifurcation. On the other hand, when
$\lambda_j$ decreases through $-1$, it becomes unstable via
period-doubling bifurcation.

It follows from the coupling condition (\ref{eq:ccd}) that
\begin{equation}
{\sum_{l=1}^{N}} V^{(l)}(X)=0.
\label{eq:sr}
\end{equation}
Hence the Lyapunov exponent $\sigma_0$ 
becomes
\begin{equation}
\sigma_0 = \displaystyle {\lim_{m \rightarrow \infty}} {1 \over m}
{\sum _{t=0}^{m-1}} \ln|u'(X^*_t)|.
\end{equation}
It is just the Lyapunov exponent of the uncoupled 1D map.
While there is no coupling effect on $\sigma_0$, coupling generally
affects the other Lyapunov exponents $\sigma_j$'s $(j \neq 0)$.

In case of the global coupling of the form (\ref{eq:gc}), the reduced
coupling functions become:
\begin{equation}
V^{(l)}(X) = \left \{  \begin{array}{cc}
                    (1-N) V(X) & {\rm for}\;l=1, \\
                     V(X) &  {\rm for}\;l \neq 1,
                       \end{array}
             \right.
\label{eq:mcrcf}
\end{equation}
where
\begin{equation}
V(X) = {c \over N} r'(X).
\end{equation}
Substituting $V^{(l)}$'s into Eq.~(\ref{eq:lexp}), we obtain:
\begin{equation}
\sigma_j = \left \{ \begin{array}{l}
\displaystyle {\lim_{m \rightarrow \infty}} {1 \over m}
{\sum _{t=0}^{m-1}} \ln|u'(X^*_t)| \;{\rm for}\; j=0, \\
\displaystyle {\lim_{m \rightarrow \infty}} {1 \over m}
{\sum _{t=0}^{m-1}} \ln|u'(X^*_t)-c r'(X^*_t)|\;{\rm for}\; j \neq 0.
                    \end{array}
         \right.
\label{eq:glexp}
\end{equation}
Here $\sigma_0$ is just the tangential Lyapunov exponent $\sigma_\|$
of Eq.~(\ref{eq:sigmal}), characterizing the mean exponential rate of 
divergence of nearby orbits along the synchronization line 
(\ref{eq:sl}). On the other hand, all the other ones $\sigma_j$ 
$(j \neq 0)$ (characterizing the mean exponential rate of divergence 
of nearby orbits across the synchronization line) are the same as
the transversal Lyapunov exponent $\sigma_\bot$ of 
Eq.~(\ref{eq:sigmac}), i.e.,
\begin{equation}
\sigma_1=\cdots=\sigma_{N-1} = \sigma_\bot.
\label{eq:sile}
\end{equation}
Consequently, there exist only one independent transversal Lyapunov 
exponent $\sigma_\bot$.

As for the case of two coupled maps in Sec.~\ref{sec:SB},
we choose $u(X) = 1 - A X^2$ as the uncoupled 1D map, and consider
a global-coupling case with $r(X)=u(X)$, i.e., a global-coupling of 
the form
\begin{equation}
v(X(1),\dots,X(N)) = { c \over N} \sum_{m=1}^{N} [u(X(m))-u(X(1))]
\;\;{\rm for}\;N \geq 2.
\label{eq:mcgc}
\end{equation}
[Here the case for $N=2$ is just the dissipative coupling of
Eq.~(\ref{eq:DC}).]
For this kind of global coupling, the transversal Lyapunov exponent
(\ref{eq:sile}) is given by
\begin{equation}
\sigma_\bot=\sigma_\|+\ln|1-c|.
\end{equation}

As in Sec.~\ref{sec:SB}, we also consider the intermittency associated 
with a saddle-node bifurcation to a pair of synchronous periodic
orbits with period $3$. Then the phase diagram in Fig.~\ref{fig:PD} 
holds for all $N$ $(N \geq 2)$ globally-coupled maps, because the 
two independent Lyapunov exponents $\sigma_\|$ and $\sigma_\bot$ 
are the same, irrespectively of $N$. Thus there exists a critical 
line segment joining two ends $c_l=0$ and $c_r=2$ on the synchronous 
saddle-node-bifurcation line $A=A_c=1.75$ in the $c-A$ plane for any 
$N$ globally-coupled case. The critical behaviors of $N$ 
globally-coupled maps for $N>2$ are also essentially the same as 
those of two dissipatively-coupled maps, in which case there are two 
kinds of critical behaviors (for details of the $N=2$ case, refer to 
Sec.~\ref{sec:SB}).

\subsection{Renormalization analysis of many globally-coupled maps}
\label{sub:RA}

As seen in Sec.~\ref{sub:LE}, there exists the same critical line 
segment associated with intermittency for $N$ $(N=2,3,4,\dots)$ 
globally-coupled maps, irrespectively of $N$. Following the same 
procedure of Sec.~\ref{sec:RA} for two coupled maps, we extend the 
renormalization results of two coupled maps to arbitrary $N$ 
globally-coupled maps. We thus find two fixed maps of the reduced 
renormalization transformation and obtain their relevant eigenvalues. 
One fixed map with no relevant CE's is associated with the critical 
behavior inside the critical line segment, while the other one with 
relevant CE's is associated with the critical behavior at both ends.

To study the intermittency in the vicinity of a saddle-node 
bifurcation to a pair of synchronous orbits with period $p$ in an 
$N$-coupled map $M$ of Eq.~(\ref{eq:mcm}), consider its $p$th 
iterate $M^{(p)}$,
\begin{eqnarray}
M^{(p)}: X_{t+1} (m) &=& W^{(p)}(\sigma^{m-1}{\bf X}_t) \nonumber \\
           &=& W^{(p)}(X_t(m),X_t(m+1),\dots,X_t(m-1)), \nonumber \\
                     &&\;\;\;\;\;\;\;\;\;\;\;\;\;\;\;m=1,\dots,N,
\label{eq:pi}
\end{eqnarray}
where $W^{(p)}$ satisfies a recurrence relation
\begin{equation}
W^{(p)}({\bf X}) = W(W^{(p-1)}({\bf X}), W^{(p-1)}(\sigma {\bf X}),
\dots, W^{(p-1)}(\sigma^{N-1} {\bf X})),
\end{equation}
and it can be also decomposed into two parts, the uncoupled part 
$u^{(p)}$ and the remaining coupling part, i.e.,
\begin{equation}
W^{(p)}({\bf X}) = u^{(p)}(X(1)) + [W^{(p)}({\bf X}) - u^{(p)}(X(1)].
\end{equation}
For the threshold value $A_c$ of a synchronous saddle-node 
bifurcation, $p$ synchronous fixed points of $M^{(p)}$ appear. 
Shifting the origin of coordinates $(X(1),\dots,X(N))$ to one of the 
$p$ fixed points $(X^*(1),\dots,X^*(N))$ $(X^*(1)=\cdots=X^*(N) 
\equiv X^*=u^{(p)} (X^*)$ for $A=A_c$), we have
\begin{eqnarray}
T: x_{t+1} (m) &=& F(\sigma^{m-1} {\bf x}_t)) \nonumber \\
               &=& f(x_t(m)) + g(x_t(m),\dots,x_t(m-1)), \nonumber \\
               &&\;\;\;\;\;\;\;\;\;\;\;\;\;\;\;m=1,\dots,N,
\label{eq:Tcm}
\end{eqnarray}
where
\begin{eqnarray}
f(x(1)) &=& u^{(p)}(x(1) + X^*) - X^*, \\
g({\bf x}) &=& W^{(p)} ( x(1)+ X^*,\dots,x(N) + X^* ) - u^{(p)}
(x(1)+X^*).
\end{eqnarray}
Here ${\bf x} = (x_1,\dots,x_N)$, the uncoupled part $f$ for the 
critical case $A=A_c$ satisfies the condition (\ref{eq:BC}), and the 
coupling function also obeys the condition (\ref{eq:ccd}).

We employ the same renormalization transformation $\cal N$ of
Eq.~(\ref{eq:RT}) with the rescaling operator $\alpha I$, where
$\alpha$ is a rescaling factor, and $I$ is the $N \times N$ identity 
matrix. Applying the renormalization operator $\cal N$ to the 
$N$-coupled map (\ref{eq:Tcm}) $n$ times, we obtain the $n$-times 
renormalizaed map $T_n$ of the form,
\begin{eqnarray}
T_n : x_{t+1}(m) &=& F_n (\sigma^{m-1} {\bf x}_t) \nonumber \\
        &=& f_n (x_t(m)) + g_n (x_t(m),\dots,x_t(m-1)), \nonumber \\
        && \;\;\;\;\;\;\;\;\;\;\;\;\;\;\;m=1,\dots,N.
\end{eqnarray}
Here the uncoupled and coupling parts $f$ and $g$ satisfy the 
following recurrence relations:
\begin{eqnarray}
f_{n+1} (x(1)) &=& \alpha f_n ( f_n ( {x(1) \over \alpha})),
\label{eq:mcft} \\
g_{n+1} ({\bf x}) &=& \alpha f_n ( F_n( {{\bf x} \over \alpha}))
 + \alpha g_n (F_n( {{\bf x} \over \alpha}), \dots,
 F_n( {{\sigma^{N-1} \bf x} \over \alpha})) \nonumber \\
&& - \alpha  f_n ( f_n( {x(1) \over \alpha})). \label{eq:mcgt}
\end{eqnarray}
Then Eqs.~(\ref{eq:mcft}) and (\ref{eq:mcgt}) define a renormalzation
operator $\cal R$ of transforming a pair of functions $(f,g)$:
\begin{equation}
 \left( \begin{array}{c}
                    {f_{n+1}} \\
                    g_{n+1}
                   \end{array}
           \right)
= {\cal R}  \left( \begin{array}{c}
                    {f_n} \\
                    g_{n}
                   \end{array}
           \right).
\end{equation}

A critical map $T_c$ is attracted to a fixed map $T^*$ under iterations 
of the renormalization transformation $\cal N$,
\begin{eqnarray}
T^* : x_{t+1}(m) &=& F^* (\sigma^{m-1} {\bf x}_t) \nonumber \\
 &=& f^*(x_t(m)) \nonumber \\
 &&+ g^*(x_t(m),x_t(m+1),\dots,x_t(m-1)), \nonumber \\
 &&\;\;\;\;\;\;\;\;\;\;\;\;\;\;\;m=1,\dots,N.
\end{eqnarray}
Here $(f^*,g^*)$ is a fixed point of the renormalization operator 
$\cal R$, i.e., $(f^*,g^*) = {\cal R} (f^*,g^*)$. Since $f^*$ is just 
the 1D fixed map (\ref{eq:1Dfm}), only the equation for the coupling 
fixed function $g^*$ is left to be solved.

As in case of two coupled maps, we construct a tractable
recurrence equation for a reduced coupling function of $g({\bf x})$
defined by
\begin{equation}
G^{(l)}(x) = \left. {{\partial g({\bf x})} \over {\partial x(l)}}
            \right|_{x(1)=\cdots=x(N)=x},
\end{equation}
because it is not easy to directly solve the equation for the coupling
fixed function. Differentiating the recurrence equation 
(\ref{eq:mcgt}) for $g$ with respect to $x(l)$ $(l=1,\dots,N)$ and 
setting $x(1)=\cdots=x(n)=x$,
we obtain:
\begin{eqnarray}
G_{n+1}^{(l)} &=& f'_n ( f_n ({x \over \alpha}) )\, G_n^{(l)}
({x \over \alpha})   \nonumber \\
&& + G_n^{(l)}( f_n ({x \over \alpha}))\, f'_n ({x \over \alpha}) 
\nonumber \\
&&+ {\sum_{i=1}^N} G_n^{(i)} ( f_n( {x \over \alpha})) \,
G_n^{(l-i+1)} ({x \over \alpha}).
\label{eq:mcrcfre}
\end{eqnarray}
The reduced coupling functions $G^{(l)}$'s satisfy the sum rule
(\ref{eq:sr}), i.e., $\displaystyle {\sum_{l=1}^{N}} G^{(l)}(x)=0$,
and they also satisfy $G^{(l)}(x) = G^{(l+N)}(x)$ due to the periodic
boundary condition.

As shown in Eq.~(\ref{eq:mcrcf}), there exists only one independent
reduced coupling function $G(x)$ for the globally-coupled case, 
such that
\begin{eqnarray}
G^{(2)}(x) &=& \cdots = G^{(N)}(x) \equiv G(x), \nonumber \\
G^{(1)}(x) &=& (1-N) G(x).
\label{eq:mcircf}
\end{eqnarray}
Then it is easy that the successive images $\{ G^{(l)}_n (x) \}$ of
$\{ G^{(l)} (x) \}$ under the transformation (\ref{eq:mcrcfre}) also
satisfy Eq.~(\ref{eq:mcircf}) [i.e., $G^{(2)}_n = \cdots = 
G^{(N)}_n(x) \equiv G_n(x),\;G^{(1)}_n(x) = (1-N) G_n(x)$].
Consequently, there remains only one recurrence equation for the 
independent reduced coupling function $G(x)$ \cite{Kim3}:
\begin{eqnarray}
G_{n+1}(x) &=& [{f'_n}({f_n}({\frac {x} {\alpha}}))- NG_{n}({f_n}
({\frac {x} {\alpha}}))] G_{n}({\frac {x} {\alpha}}) \nonumber \\
&& + G_{n}({f_n}({\frac {x} {\alpha}})) {f'_n}({\frac {x} {\alpha}}).
\label{eq:mcRRE}
\end{eqnarray}
Then, together with Eq.~(\ref{eq:mcft}), Eq.~(\ref{eq:mcRRE})
defines a reduced renormalization operator $\tilde{\cal R}$
of transforming a pair of functions $(f,G)$:
\begin{equation}
 \left( \begin{array}{c}
                    {f_{n+1}} \\
                    G_{n+1}
                   \end{array}
           \right)
= {\tilde{\cal R}}  \left( \begin{array}{c}
                    {f_n} \\
                    G_{n}
                   \end{array}
           \right).
\label{eq:mcRRO}
\end{equation}
Since the reduced renormalization transformation (\ref{eq:mcRRO})
holds for any globally-coupled cases of $N \geq 2$, it can be regarded 
as a generalized version of Eq.~(\ref{eq:RRO}) for the case of two 
coupled maps.

A pair of critical functions $(f_c,G)$ is attracted to a pair of fixed
functions $(f^*,G^*)$ under iterations of $\tilde {\cal R}$.
Here $f^*$ is the 1D fixed map (\ref{eq:1Dfm}), and $G^*(x)$ is the
independent reduced coupling fixed-function of $g^*({\bf x})$ [i.e.,
$G^{*(2)}(x)=\cdots=G^{*(N)}(x)=G^*(x),\;G^{*(1)}(x)= (1-N) G^*(x)$].
As in the two-coupled case $(N=2)$ of Eqs.~(\ref{eq:SS1}) and 
(\ref{eq:SS2}), we find two series solutions for $G^*$:
\begin{eqnarray}
G^*_I(x) &=& {1 \over N} [1 + z a x^{z-1} + z (z-{1 \over 2}) a^2 
x^{2(z-1)} + \cdots], \label{eq:mcSS1} \\
G^*_{II}(x) &=& {1 \over N} [b a x^{z-1} + b ({{3z-b} \over 2}-
{1 \over 2}) a^2 x^{2(z-1)} + \cdots].
\label{eq:mcSS2}
\end{eqnarray}
Here $a$ and $b$ are arbitrary constants.
The solutions for $G^*$ have a common factor $1 \over N$, and hence 
the function $N\,G^*(x)$ becomes the same, independently of $N$; this 
can be also easily understood by looking at the structure of 
Eq.~(\ref{eq:mcRRE}). In case of $G^*_I(x)$, we can sum the series and 
obtain a closed-form solution,
\begin{equation}
G^*_I(x) = {1 \over N} f'^*(x).
\end{equation}
However, unfortunately we cannot sum the series in $G^*_{II}(x)$
except for the cases $b=0$ and $z$ where we obtain closed-form
solutions,
\begin{equation}
G^*_{II} (x) = \left \{
                  \begin{array}{c}
                  0 \;\;\;\;\;\;{\rm for}\;b=0, \\
                  {1 \over N} [{f^*}'(x)-1]\;\;\;{\rm for}\;b=z.
                  \end{array}
          \right.
\end{equation}

Linearizing the renormalization transformation $\tilde{\cal R}$ at its 
fixed point $(f^*,G^*)$, we obtain a reduced linearized operator 
$\tilde{\cal L}$ transforming a pair of infinitesimal perturbations 
$(h,\Phi)$:
\widetext
\begin{equation}
 \left( \begin{array}{c}
                    {h_{n+1}} \\
                    {\Phi _{n+1}}
                   \end{array}
           \right)
= {\tilde{\cal L}}  \left( \begin{array}{c}
                     {h_n} \\
                     {\Phi_n}
                           \end{array}
                    \right)
= \left( \begin{array}{cc}
           {\tilde{\cal L}_1} & \;\;\; 0\\
           {\tilde{\cal L}_3} & \;\;\; {\tilde{\cal L}_2}
          \end{array}
           \right) \;
           \left( \begin{array}{c}
                    {h_n} \\
                    {\Phi_n}
                   \end{array}
              \right),
\end{equation}
where
\begin{eqnarray}
&{h_{n+1}}(x)&= [{{\tilde{\cal L}}_1} {h_n}](x) =
{\alpha}{f^*}'({f^*}(\frac{x}{\alpha})) {h_n}(\frac{x}{\alpha}) +
{\alpha} {h_n}({f^*}(\frac{x}{\alpha})), \label{eq:mcREh} \\
&{{\Phi}_{n+1}}(x) &= [{{\tilde{\cal L}}_2}{\Phi _n}](x)
      +[{{\tilde{\cal L}}_3}{h_n}](x), \label{eq:mcREPhi} \\
&[{{\tilde{\cal L}}_2}{\Phi_n}](x)&=
 [{f^*}'({f^*}(\frac{x}{\alpha}))-N{G^*}({f^*}(\frac{x}{\alpha}))]
{\Phi _n}(\frac{x}{\alpha})
+ [{f^*}'(\frac{x}{\alpha})-N{G^*}(\frac{x}{\alpha})] {\Phi _n}({f^*}
(\frac{x}{\alpha})),  \\
&[{{\tilde{\cal L}}_3}{h_n}](x)&=
 [{f^*}''({f^*}(\frac{x}{\alpha})){G^*}(\frac{x}{\alpha})-N{G^*}'
 ({f^*} (\frac{x}{\alpha})){G^*}(\frac{x}{\alpha}) +{G^*}'({f^*}
 (\frac{x}{\alpha})) {f^*}'(\frac{x}{\alpha})] {h_n}(\frac{x}{\alpha}) 
 \nonumber \\
 &&\;\;\; +{G^*}(\frac{x}{\alpha}){h'_n}({f^*}(\frac{x}{\alpha})) + 
 {G^*}({f^*}(\frac{x}
{\alpha})){h'_n}(\frac{x}{\alpha}).
\end{eqnarray}
\narrowtext
It follows from the reducibility of $\tilde{\cal L}$ into a semiblock 
form that to determine the eigenvalues of $\tilde{\cal L}$ it is 
sufficient to solve the eigenvalue problems for ${\tilde{\cal L}}_1$ 
and ${\tilde{\cal L}}_2$ independently. Then the eigenvalues of both
${\tilde{\cal L}}_1$ and ${\tilde{\cal L}}_2$ give the whole spectrum
$\tilde{\cal L}$.

The eigenvalue equation for ${\tilde{\cal L}}_1$ is given by
Eq.~(\ref{eq:EVEh}). As mentioned there, that is just the eigenvalue
equation for the 1D map, in which case the complete spectrum of 
eigenvalues and the corresponding eigenfunctions have been found in 
Refs.~\cite{HH}. 

We next consider an infinitesimal coupling-perturbation of the form
$(0,\Phi)$ to a fixed point $(f^*,G^*)$. If an independent reduced 
coupling perturbation $\Phi^*$ satisfies
\begin{eqnarray}
\lambda {\Phi^*}(x) &=& [{{\tilde{\cal L}}_2}{\Phi^*}](x) \nonumber \\
      &=&
 [{f^*}'({f^*}(\frac{x}{\alpha}))-N{G^*}({f^*}(\frac{x}{\alpha}))]
{\Phi^*}(\frac{x}{\alpha}) \nonumber \\
&&+ [{f^*}'(\frac{x}{\alpha})-N{G^*}(\frac{x}{\alpha})] {\Phi^*}({f^*}
(\frac{x}{\alpha})),
\label{eq:mcCE}
\end{eqnarray}
then it is called the independent reduced coupling eigenfunction with 
CE $\lambda$. Note that the eigenvalue equation (\ref{eq:mcCE}) for 
any $N$ becomes the same as that for the two-coupled case $(N=2)$, 
because the function $N G^*(x)$ is the same, irrespectively of $N$. 
Hence we follow the same procedure in Sec.~\ref{sec:RA} for two 
coupled maps, and find the same relevant CE's for any $N$ 
globally-coupled maps as follows (for more details, refer to 
Sec.~\ref{sec:RA}):
\begin{eqnarray}
&(1)& \;G^*(x) = G^*_I(x) \nonumber \\
   &&\;\;{\rm There\;exist\;no\; relevant\; CE's}. \\
&(2)& \;G^*(x) = G^*_{II}(x) \nonumber \\
&&\;\; {\rm There\;exist}\;(z-1)\;{\rm relevant\;CE's\;such\;that}\;
        \lambda_k=2^{(z-1-k) / (z-1)} \nonumber \\
 &&\;\;{\rm with\;eigenfunction}\; \Phi^*_k(x)\; {\rm of\;
  Eq.}~(\ref{eq:CEF})\;  (k=0,\dots,z-2).
\end{eqnarray}

As in Sec.~\ref{sub:LE}, we also consider the global coupling of
Eq.~(\ref{eq:mcgc}) with $r(X)=u(X)=1-AX^2$ and study the 
intermittency associated with a saddle-node bifurcation to a pair of
synchronous orbits with period $p=3$. Considering the third iterate 
$M^{(3)}$ of $M$ [see Eq.~(\ref{eq:pi})] and then shifting the 
origin of coordinates to one of the three synchronous fixed points 
for $A=A_c$, we obtain a map of the form $T$ (\ref{eq:Tcm}).
The uncoupled part $f$ has the form (\ref{eq:up}), ans hence that
corresponds to the most common case with the tangency-order $z=2$. 
The independent reduced coupling function of the coupling part 
$g({\bf x})$ is also given by
\begin{equation}
G(x) = {e \over N} f'(x),\; e=c^3-3c^2+3c.
\end{equation}

Consider a pair of initial functions $(f_c,G)$ on the synchronous
saddle-node bifurcation line $A=A_c$, where $f_c(x)$ is just the 1D 
critical map and $G(x)={e \over N} \,f'_c(x)$. By successive 
applications of the reduced renormalization operator 
$\tilde {\cal R}$ of Eq.~(\ref{eq:mcRRO}) to $(f_c,G)$, we have
\begin{eqnarray}
&f_n&(x) = \alpha f_{n-1}(f_{n-1}({x \over \alpha})),\;\;
G_n(x)= {e_n \over N}\,f'_n(x) \\
&e_n& = 2e_{n-1} -  e^2_{n-1}, \label{eq:mcred}
\end{eqnarray}
where the rescaling factor for $z=2$ is $\alpha=2$, $f_0(x) = f_c(x)$,
$G_0(x)=G(x)$, and $e_0=e$.
Here $f_n$ converge to the 1D fixed function $f^*(x)$ of 
Eq.~(\ref{eq:1Dfm}) with $z=2$ as $n \rightarrow \infty$.

As shown in Sec.~\ref{sec:SB}, any initial $e$ inside the open 
interval $(0,2)$ converges to the superstable fixed point $e^*=1$ 
under successive iterations of the transformation (\ref{eq:mcred}). 
The left end of the interval is un unstable fixed point $e^*=0$, 
which is also the image of the right end $e=2$ under the 
transformation (\ref{eq:mcred}); all the other points outside the 
interval diverges to the minus infty under iterations of the 
transformation (\ref{eq:mcred}). One can see easily that the 
interval $[0,2]$ of the parameter $e$ corresopnds to a critical line 
segment joining two ends $c_l=0$ and $c_r=2$ on the synchronous 
saddle-node bifurcation line $A=A_c$ in the $c-A$ plane.
Hence any initial $G(x)$ inside the critical line segment is 
attracted to the first independent reduced coupling fixed function
$G^*_I(x)= {1 \over N} f^*(x)$ under iterations of $\tilde {\cal R}$, 
while $G(x)'s$ at both ends are attracted to the second independent 
reduced coupling fixed function $G^*_{II}(x) =0$ with $b=0$.

\section{Summary}
\label{sec:SUM}
The critical behaviors for intermittency in two coupled 1D maps are
studied by a renormalization method. We find two fixed maps of the
renormalization transformation. Although they have common relevant 
eigenvalues associated with scaling of the control parameter of the 
uncoupled 1D map, their relevant CE's associated with coupling
perturbations vary depending on the fixed maps. These two fixed maps 
are also found to be associated with the critical behavior near a 
critical line segment. One fixed map with relevant CE's govern
the critical behavior at both ends of the critical line segment, 
while the other one with no relevant CE's governs the critical 
behavior inside the critical line segment. We also extend the results 
of the two coupled 1D maps to many globally-coupled 1D maps.

\acknowledgments
This work was supported by the the Korea Research Foundation under 
Project No. 1997-001-D00099.

\begin{figure}
\caption{
Phase diagram for a dissipative-coupling case (4.2). Here solid
circles denote the data points on the $\sigma_\bot =0$ curve. The 
region enclosed by the $\sigma_\bot =0$ curve is divided into two 
parts dnoted by $P$ and $C$. A synchronous period-3 (chaotic) 
attractor with $\sigma_\| <0$ $(\sigma_\|>0)$ exists in the part 
$P$ $(C)$. The boundary curve denoted by a solid line between the 
$P$ and $C$ parts is just the critical line segment.
}
\label{fig:PD}
\end{figure}

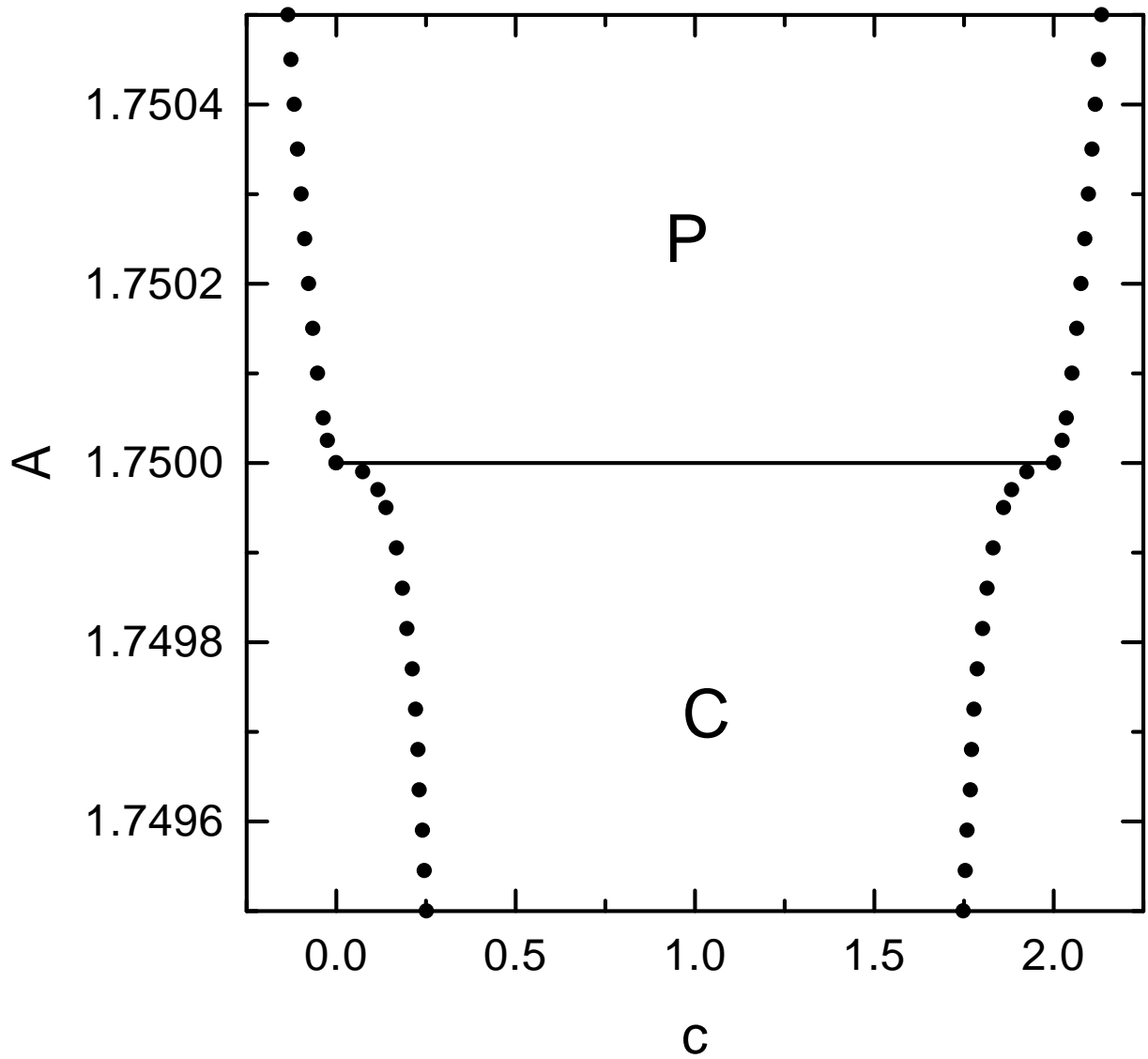
\end{document}